\documentclass[aps,pra,twocolumn,groupedaddress,showpacs]{revtex4}

\usepackage{amsmath} 
\usepackage{amssymb} 
\usepackage{graphicx}

\newcommand{\op}[1]{\hat{\rm{#1}}}

\begin{document}


\title{Ionization behavior of molecular hydrogen in intense laser fields:\\
  Influence of molecular vibration and alignment}

\author{Johann F\"orster}

\author{Yulian V.~Vanne}

\author{Alejandro Saenz}

\affiliation{ AG Moderne Optik, Institut f\"ur Physik, Humboldt-Universit\"at
  zu Berlin, Newtonstra{\ss}e 15, D\,-\,12\,489 Berlin, Germany }

\date{\today}

\begin{abstract}
  The alignment- and internuclear-distance dependent ionization of
  H$_2$ exposed to intense, ultrashort laser fields is studied by
  solving the time-dependent two-electron Schr\"odinger equation. In
  the regime of perturbative few-photon ionization, a strong
  dependence of the ionization yield on the internuclear distance is
  found.  While this finding confirms a previously reported breakdown
  of the fixed-nuclei approximation for parallel alignment, a simpler
  explanation is provided and it is demonstrated that this breakdown
  is not due to vibrational dynamics during the laser pulse. The
  persistence of this effect even for randomly aligned molecules is
  demonstrated.  Furthermore, the transition from the multiphoton to
  the quasi-static (tunneling) regime is investigated considering
  intense 800\,nm laser pulses.  While the obtained ionization yields
  differ significantly from the prediction of Ammosov-Delone-Krainov
  rates, we find a surprisingly good quantitative agreement after
  introducing a simple frequency-dependent correction to the standard
  tunneling formula.
\end{abstract}


\maketitle

%
\section{Introduction}
%

The rapid development of intense, ultrashort laser pulses during the
past decade offers the prospects to measure and manipulate molecules
on their natural time scales (few femtoseconds to attoseconds).  By
investigating the response of small molecules to these laser fields,
concepts to produce a real-time movie of the electronic and nuclear
dynamics triggered in these molecules have been developed.  The
high-harmonic radiation emitted from these molecules contains
information which may be used for, e.g., orbital tomography
\cite{sfm:itat04}, probing of nuclear dynamics with sub-fs resolution
\cite{sfm:lein05,sfm:bake06,sfm:farr11a,sfm:krau13,sfm:foer13}, and
following coupled electron-nuclear dynamics with time-resolved
high-harmonic spectroscopy \cite{sfm:worn10c, sfm:woer11}.
Noteworthy, already the electrons emitted by ionization (seen as the
first step of high-harmonic generation) contain structural information
suitable for orbital imaging \cite{sfm:meck08, sfm:petr10a}.
Moreover, a process termed Lochfra\ss\ allows to create nuclear wave
packets in neutral molecules and to measure them with sub-femtosecond
and sub-\AA ngstr\"om resolution adopting a pump-probe scheme
\cite{sfm:goll06, sfm:ergl06, sfm:fang08a}.

In view of the many promising proposals a deeper understanding of the
molecular response to intense, ultrashort laser fields is desirable.
Compared to atoms, the nuclear degrees of freedom (vibration and
rotation) as well as the multi-centered (non-spherically symmetric)
electronic structure of molecules increase the complexity regarding
their theoretical treatment.  Thus, even molecular hydrogen H$_2$,
despite being the simplest neutral molecule, remains a great challenge
for theory when exposed to intense laser fields.  This is especially
true, if the correlated two-electron Schr\"odinger equation is solved
in all six dimensions. In the case of large laser frequencies, low
intensities, and not too extremely short laser-pulse durations,
lowest-order perturbation theory (LOPT) may be used.  Thus, at first,
perturbative one-photon ionization \cite{dia:mart99} (and references
therein) and later on two- to four-photon ionization \cite{sfm:apal02}
of H$_2$ have been studied.

The direct numerical solution of the time-dependent Schr\"odinger
equation (TDSE) describing H$_2$ in intense laser fields for fixed
nuclei and a parallel alignment was first realized on a sophisticated
grid \cite{sfm:haru00} and then using a configuration-interaction
expansion built from H$_2^+$ orbitals expressed in prolate spheroidal
coordinates \cite{sfm:awas05}.  In the perturbative regime, good
quantitative agreement between LOPT and TDSE ionization yields has
been found \cite{sfm:awas05}.  This latter approach has also been
applied, e.g., for longer wavelengths and higher intensities as well
as for non-parallel orientations of the laser polarization with
respect to the molecular axis
\cite{sfm:awas06,sfm:vann08,sfm:vann09,sfm:vann10}.  It was shown in
\cite{sfm:vann10} that a simplified treatment using the molecular
strong-field approximation (in velocity gauge) can contradict the
behavior obtained from the direct TDSE solution even qualitatively.

In a different approach based on an expansion in Born-Oppenheimer
eigenstates and a single-center expansion for the electronic problem,
the TDSE has been solved accounting also for vibrational dynamics
\cite{sfm:pala06} (neglecting non-adiabatic couplings).  Large
differences between the treatment that included the vibrational
dynamics and the fixed-nuclei approximation were found
\cite{sfm:pala06, sfm:pala07, sfm:sanz07}.  Later applications of this
approach concentrated mainly on low laser intensities and studied,
e.g., the decay of autoionizing states \cite{sfm:rivi12, sfm:silv12}.
Another, more recently introduced approach, again using prolate
spheroidal coordinates but Laguerre and Legendre polynomials as basis
functions, has been applied to investigate enhanced ionization
occurring at large internuclear distances \cite{sfm:dehg10}.

In the following section, the method to solve the two-electron TDSE
and the basis-set parameters used are briefly discussed.  In
Section~\ref{sec:Perturbative}, the method is applied in the
perturbative multiphoton regime and compared to the results in
Refs.~\cite{sfm:pala06, sfm:pala07, sfm:sanz07}.  In particular, the
breakdown of the widely used fixed-nuclei approximation is
re-investigated in detail with the present approach. Furthermore, the
study is extended to non-parallel (random) orientations of the laser
polarization with respect to the molecular axis.  The
internuclear-distance dependent ionization behavior in the transition
from the multiphoton regime to the quasi-static regime is studied in
Section~\ref{sec:800nm} for the widely adopted Ti:sapphire wavelength.
A parallel as well as a perpendicular orientation of the laser
polarization with respect to the molecular axis is considered.  The
ionization yield is compared to the one obtained from the approximate
Ammosov-Delone-Krainov (ADK) tunneling rates \cite{sfa:ammo86}.
Returning to the original Perelomov-Popov-Terent'ev \cite{sfa:pere66}
theory, a correction to the ADK tunneling rate is introduced and
compared to the TDSE results. Unless noted otherwise, atomic units
with $\hbar=e=m_e=4\pi \epsilon_0 = 1$ are adopted in this work.

%
\section{Method}\label{sec:Method}
%

The method to solve the TDSE describing molecular hydrogen exposed to
a laser field within the fixed-nuclei approximation is discussed in
detail in previous works~\cite{sfm:awas05,sfm:vann08,sfm:vann09}.
Briefly, the TDSE
\begin{eqnarray}
i \frac{\partial}{\partial t} \psi(\mathbf{r},t) = \left(\op{H}_0 +
\op{V}(t)\right) \psi(\mathbf{r},t)
\end{eqnarray}
is solved by expanding the time-dependent electronic wave function
$\psi(\mathbf{r},t)$ in terms of eigenstates of the field-free
electronic Hamiltonian $\op{H}_0$ ($\mathbf{r}$ represents the set of
both electronic coordinates).  The latter eigenstates are obtained
from a configuration-interaction (CI) calculation in which the Slater
determinants are formed with the aid of H$_2^+$ eigenstates expressed
in terms of $B$ splines in prolate spheroidal
coordinates~\cite{dia:vann04}.  The linearly polarized laser pulse is
described classically by the vector potential
\begin{eqnarray}
  \vec{A}(t) = \left\{\begin{array}{cl} \vec{A}_0 \cos^2(\pi t/T) \sin(\omega t
      + \varphi) & \mbox{for }|t|\leq T/2\\ 0 & \mbox{elsewhere} \end{array}\right.
\end{eqnarray}
with laser frequency $\omega$, total pulse duration $T=2\pi
n_{\mathrm{c}}/\omega$ (number of cycles $ n_{\mathrm{c}}$), and
carrier-envelope phase $\varphi$.  The interaction potential reads
$\op{V}(t) = \hat{\mathbf{p}} \cdot \vec{A}(t)$ (dipole approximation,
velocity gauge).  To obtain the ionization yield, the electronic
problem is solved for a single fixed internuclear distance $R$ and
alignment angle $\theta$.  Here, $\theta = 0$ corresponds to a
parallel alignment ($\parallel$) of the polarization direction with
respect to the molecular axis and $\theta = \pi/2$ corresponds to a
perpendicular alignment ($\perp$).  The ionization yield $Y(R,\theta)$
is then given by the population of all electronic continuum states
after the laser pulse.

The main basis-set parameters adopted for the present calculations are
discussed in detail in~\cite{sfm:vann09}.  Briefly, a box size of
about $350$ a.u.\ with 350 B~splines of order $k=10$ were used along
the $\xi$ coordinate (knot distribution: geometric progression with
$g=1.05$ for the first 40 intervals and linear progression
afterwards).  30 B~splines of order 8 with a linear knot sequence were
used along the $\eta$ coordinate and highly oscillatory H$_2^+$
orbitals with more than 19 nodes along $\eta$ were omitted in the CI
expansion. The CI expansion consists of a very long configuration
series where one electron occupies the H$_2^+$ ground-state
$1\sigma_g$ while the other is occupying one of the remaining (bound
or discretized continuum) H$_2^+$ eigenstates. Together with
additional CI configurations which represent doubly excited situations
(responsible for the description of correlation and doubly excited
states), this corresponds to about 6000 configurations per symmetry.
Obtaining sufficiently converged TDSE solutions is computationally
much more challenging for the $800$\ nm laser pulses discussed in
Sec.~\ref{sec:800nm} compared to the perturbative regime in
Sec.~\ref{sec:Perturbative}.

For the results shown in Sec.~\ref{sec:Perturbative}, states with
energies up to $1$ a.u.\ above the ionization threshold were included
in the time propagation ($0.5$ a.u.\ are already sufficient). For
non-parallel alignments, states with maximal absolute values of the
component of the total angular momentum along the internuclear axis
were included up to $\Lambda_{\textrm{max}}=7$ (convergence was found
already with $\Lambda_{\textrm{max}}=4$). Furthermore, convergence
with respect to the box size has been checked by doubling the box
size.  Most importantly, the CI expansion as described in
\cite{sfm:vann09} leads to a non-perfect description of the
ground-state of H$_2$, especially its energy. This manifests in a
shifted frequency $\omega = \omega_{\mathrm{num}} + \Delta\omega$ with
$\Delta\omega = 0.0092$ a.u.\ which is used for our results shown in
Sec.~\ref{sec:Perturbative} ($\omega_{\mathrm{num}}$ is the frequency
used in the numerical calculation). It was found that this frequency
shift $\Delta\omega$ is reduced when using a more complete CI
expansion, but other than this shift no significant change in the
ionization behavior was observed.

\begin{figure}
\begin{center}
     \includegraphics[width=0.45\textwidth]{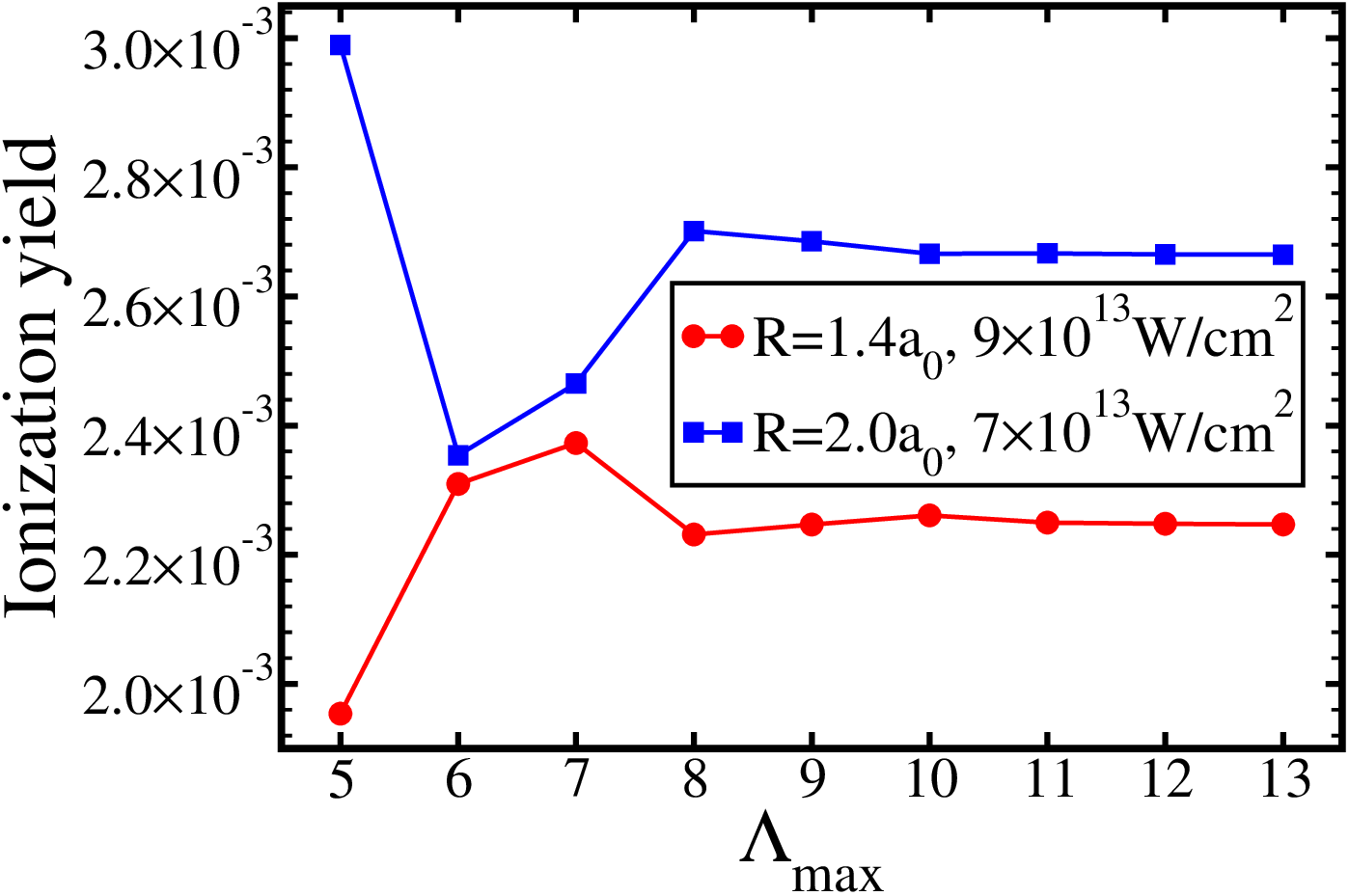}
     \caption{\label{fig:Lmax_conv} (Color online) Convergence of the
       total ionization yield for a perpendicular-aligned H$_2$
       molecule with respect to the maximal absolute value of the
       component of the total angular momentum along the internuclear
       axis, $\Lambda_{\textrm{max}}$, in a 20-cycle cos$^2$-shaped
       800\,nm laser pulse.}
\end{center}
\end{figure}

In the case of $800$ nm (Sec.~\ref{sec:800nm}), states with energies
up to $10$ a.u.\ above the ionization threshold were included as in
\cite{sfm:vann09}. Furthermore, at this wavelength the treatment of
the perpendicular alignment is much more challenging than the parallel
one. Fig.~\ref{fig:Lmax_conv} shows the typical convergence of the
total ionization yield with respect to $\Lambda_{\textrm{max}}$ for
two different intensities and internuclear distances.  For the purpose
of this work, the values of $\Lambda_{\textrm{max}}=7-11$ were used to
obtain sufficiently converged ionization yields, while higher values
would be needed to obtain fully converged photoelectron spectra. A
similar convergence behavior was observed with respect to the box
size.

%

Within the {\it fixed-nuclei approximation} (FNA), the ionization
yield $Y_{\mathrm{FNA}}(\theta)=Y(R_{\mathrm{eq}},\theta)$ is
approximated by the electronic response at the equilibrium
internuclear distance $R_{\mathrm{eq}}=1.4$ a.u.  While a treatment
fully including vibrational dynamics (FULL) as in \cite{sfm:pala06,
  sfm:pala07,sfm:sanz07} is beyond the scope of the present paper, we
may take nuclear vibration into account by "freezing" the initial
nuclear wave function $\chi(R)$ (vibrational ground state of the
electronic Born-Oppenheimer ground-state potential) during the
time-propagation.  Within this {\it frozen-nuclei approximation}
(FROZ) \footnote{There exists no unified terminology for this level of
  approximation. It was termed frozen-nuclei limit in
  \cite{sfm:saen02a} and we use "frozen-nuclei approximation"
  throughout this paper.}, the $R$-integrated ionization yield
\begin{eqnarray}
Y_{\mathrm{FROZ}}(\theta) = \int \mathrm{d}R\ Y(R,\theta)
\left|\chi(R)\right|^2 \label{eq:YieldRint}
\end{eqnarray}
is obtained from $Y(R,\theta)$ for a range of internuclear distances
where the nuclear wave function $\chi(R)$ of the initial state is
nonvanishing, namely $R=1.0-2.5$ a.u.\ In Sec.~\ref{sec:Perturbative}
(\ref{sec:800nm}), 61 (31) points separated by $\Delta R = 0.025$
a.u.\ ($0.05$ a.u.) were used.  Furthermore, we consider different
alignments of the laser polarization with respect to the molecular
axis, especially also non-parallel ones ($\theta \neq 0$). In the case
of a random alignment, the alignment-averaged ionization yield
\begin{eqnarray}
  Y_{\mathrm{avg,X}} = \int\limits_{0}^{\pi/2}
  \mathrm{d}\theta \sin(\theta) Y_{\mathrm{X}}(\theta)\ \mbox{with X = FNA or FROZ} \nonumber\\ \label{eq:Yieldthetaint}
\end{eqnarray}
is calculated from the fixed- or the frozen-nuclei ionization yields
$Y_{\mathrm{X}}(\theta)$ obtained for various alignment angles
$\theta$. In the case of perturbative one-photon ionization,
Eq.~(\ref{eq:Yieldthetaint}) can be simplified to
$Y_{\mathrm{avg}}^{(1\omega)} = \frac{1}{3} Y_{ \parallel} +
\frac{2}{3} Y_{\perp}$. For the here discussed few- to many-photon
ionization processes, however, the whole integration in
Eq.~(\ref{eq:Yieldthetaint}) has to be performed. Thus, 10 angles
separated by $\Delta \theta = \pi/18$ were used in
Sec.~\ref{sec:Perturbative}. Clearly, non-parallel alignments are
geometrically preferred over parallel ones due to the
\mbox{$\sin(\theta)$-factor}.  However, enhanced ionization of the
formed H$_2^+$ ion at larger internuclear distances which occurs for a
parallel alignment may obscure this fact in experiments.

%
\section{Perturbative Multiphoton Regime}\label{sec:Perturbative}
%
We study the ionization behavior for laser pulses in the perturbative
multiphoton regime with a peak intensity of $I=10^{12}$~Wcm${}^{-2}$,
total duration $T=10$ fs, carrier-envelope phase $\varphi = \pi / 2$
and laser frequencies varying between $\omega = 0.16 - 0.5$ a.u. This
allows to directly compare the results obtained with the present
method to those previously reported in \cite{sfm:pala06, sfm:pala07,
  sfm:sanz07}.

\begin{figure*}
\begin{center}
     \includegraphics[width=0.99\textwidth]{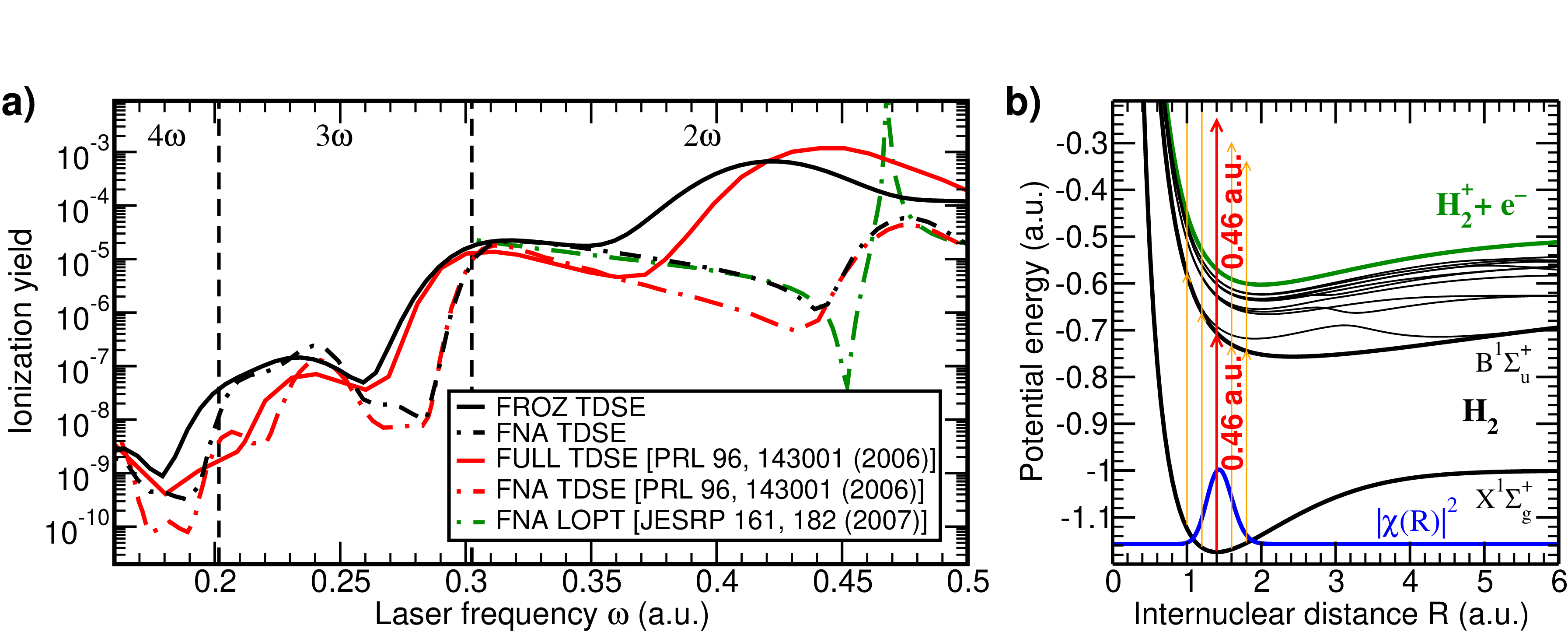}
     \caption{\label{fig:ComparisonFM} (Color online) a) Ionization
       yields as a function of the laser frequency $\omega$ for
       parallel-aligned H$_2$ exposed to $T=10$ fs, $I=10^{12}$
       Wcm${}^{-2}$ $\cos^2$-shaped laser pulses. The dashed vertical
       lines indicate the borders between the two-, the three-, and
       the four-photon ionization regimes ($2\omega$, $3\omega$,
       $4\omega$).  The ionization yields obtained within the
       fixed-nuclei (FNA TDSE) and the frozen-nuclei (FROZ TDSE)
       approximation are compared to the perturbative fixed-nuclei
       (FNA LOPT) and the TDSE results fully including vibrational
       motion (FULL TDSE) extracted from Refs.~\cite{sfm:pala06,
         sfm:pala07, sfm:sanz07}. b) Potential-energy surfaces of
       H$_2$ (black lines), H$_2^+$ ionization threshold (green line)
       and vibrational ground-state density $\left|\chi(R)\right|^2$
       (blue line). Furthermore, the resonant enhanced multiphoton
       ionization (REMPI) process ${X}^{1}\Sigma_g^+\ \rightarrow \
       {B}^{1}\Sigma_u^+\ \rightarrow \mathrm{H}_2^+(1\sigma_g) + e^-$
       for different fixed internuclear separations is indicated by
       red and orange arrows.}
\end{center}
\end{figure*}

Fig.~\ref{fig:ComparisonFM}a shows the obtained ionization yields in
direct comparison to the TDSE and LOPT results reported in
\cite{sfm:pala06, sfm:pala07, sfm:sanz07} (parallel alignment,
i.e. $\theta = 0$). We find a good qualitative and partly also
quantitative agreement between our results and the FNA TDSE results
reported in Refs.~\cite{sfm:pala06,sfm:pala07,
  sfm:sanz07}. Furthermore, the agreement between our FNA TDSE
ionization yields and the FNA LOPT yields for two-photon ionization
obtained in \cite{sfm:sanz07} is very good.  Differences between the
TDSE and LOPT ionization yields are found only around $\omega \approx
0.46$ a.u.\ where the simple LOPT approach used in \cite{sfm:sanz07}
diverges due to the resonant enhanced multiphoton ionization (REMPI)
process ${X}^{1}\Sigma_g^+\ \rightarrow \ {B}^{1}\Sigma_u^+\
\rightarrow \mathrm{H}_2^+(1\sigma_g) + e^-$ (see
Fig.~\ref{fig:ComparisonFM}b).  In fact, a similarly good quantitative
agreement between LOPT and TDSE ionization yields obtained with the
present approach was found already earlier for one-, two-, three- and
four-photon ionization of H$_2$ (pulse parameters $T=15$~fs,
$I=2\times 10^{12}$~Wcm${}^{-2}$), see Fig.~3 in
Ref.~\cite{sfm:awas05}.

Although vibrational dynamics on a timescale of the order of $10$ fs
is expected to affect the ionization behavior and thus a perfect
quantitative agreement between FROZ and FULL TDSE is not expected,
Fig.~\ref{fig:ComparisonFM}a shows that the ionization yield obtained
within FROZ TDSE behaves qualitatively surprisingly similar to the
FULL TDSE results of Refs.~\cite{sfm:pala06, sfm:pala07, sfm:sanz07}.
When comparing the FROZ and FULL TDSE ionization yields with their
respective FNA results, both treatments show the same breakdown of the
FNA, in particular an up to 3 orders of magnitude change of the
ionization yield around $\omega \approx 0.44$ a.u. Thus, already the
rather simple FROZ treatment allows for an explanation of this
preeminent breakdown of the FNA.  Clearly, the ionization yield $Y(R)$
must strongly depend on the internuclear distance $R$ in order to
obtain an ionization yield significantly different compared to the
FNA, see Eq.~(\ref{eq:YieldRint}).  At first glance, a strong
dependence of the ionization yield on the internuclear distance $R$
may not be expected since the transition dipoles do not dramatically
depend on $R$ in the vicinity of the equilibrium distance
$R_{\mathrm{eq}}=1.4$ a.u.\ (see, e.g., \cite{sfm:pala07}).  However,
for REMPI, the energy differences between the electronic states
determine where the resonance frequency or energy is located and thus
play a crucial role.  For molecules, these resonance frequencies
depend significantly on the nuclear configuration. As an example, for
the already mentioned REMPI process ${X}^{1}\Sigma_g^+\ \rightarrow \
{B}^{1}\Sigma_u^+\ \rightarrow \mathrm{H}_2^+(1\sigma_g) + e^-$,
Fig.~\ref{fig:ComparisonFM}b illustrates that at larger internuclear
distances $R>1.4$ a.u.\ significantly lower laser frequencies
$\omega<0.46$ a.u.\ are required to fulfill the resonance condition.

\begin{figure}
\begin{center}
  \includegraphics[width=0.49\textwidth]{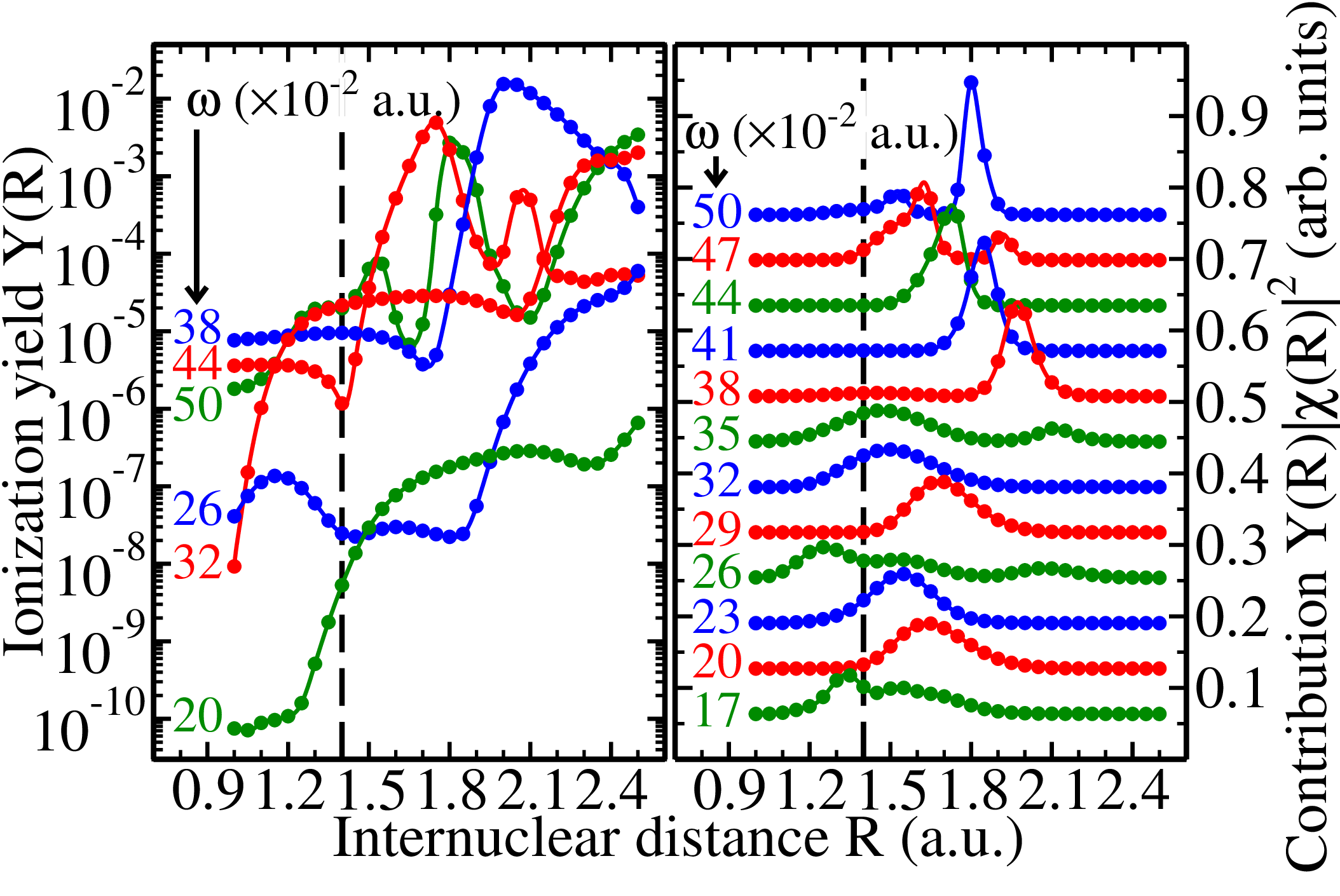}
  \caption{\label{fig:YieldDependencePerturb} (Color online)
    Ionization of parallel-aligned H$_2$ for $T=10$ fs, $I=10^{12}$
    Wcm${}^{-2}$ $\cos^2$-shaped laser pulses and different laser
    frequencies $\omega$.  The left panel shows the fixed-nuclei
    ionization yields $Y(R)$ as a function of the internuclear
    distance $R$, whereas the right panel displays the contribution
    $Y(R)\left|\chi(R)\right|^2$ to the frozen-nuclei ionization yield
    in Eq.~(\ref{eq:YieldRint}) (for better visibility scaled and
    vertically shifted).\\ }
       
      \includegraphics[width=0.49\textwidth]{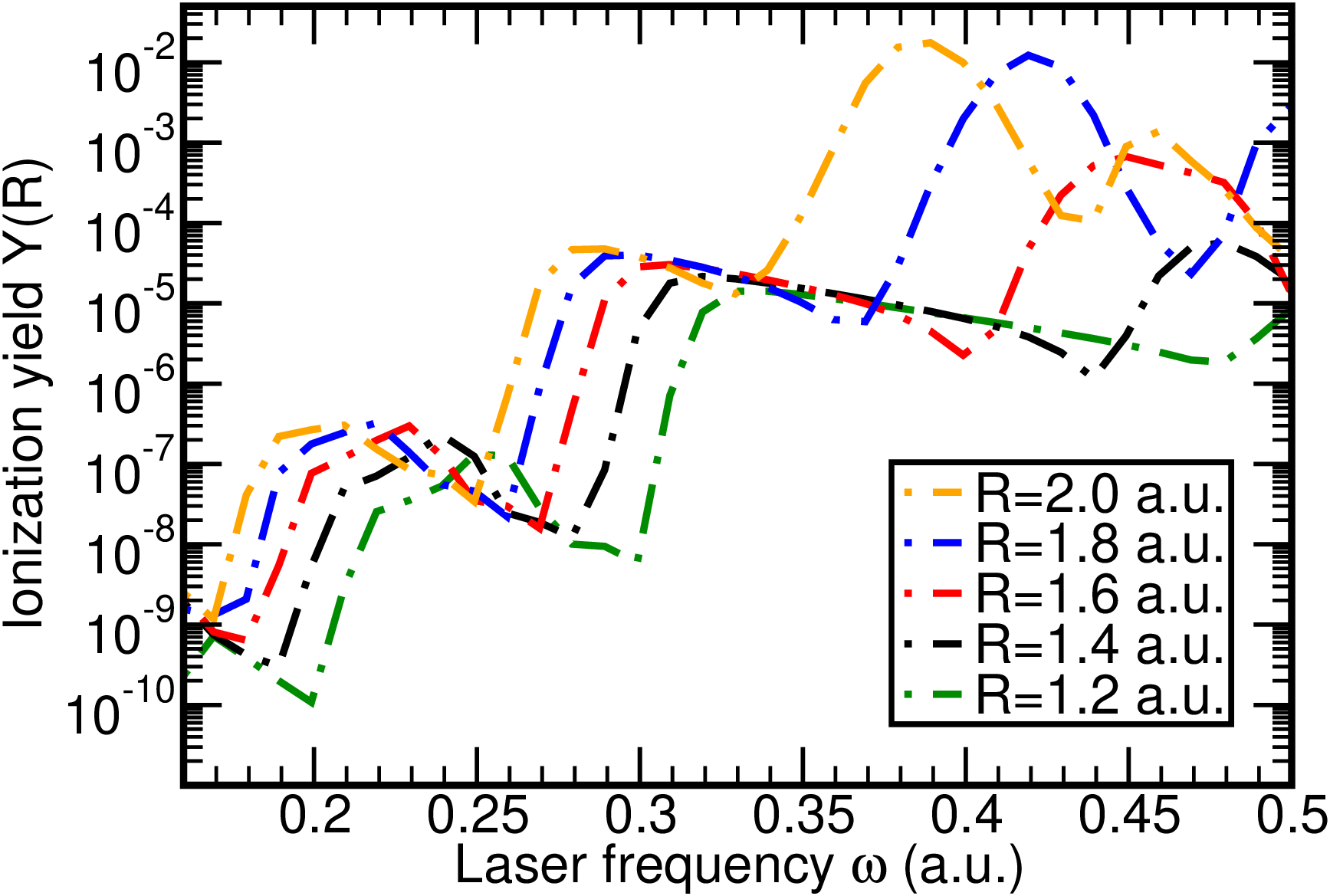}
      \caption{\label{fig:YieldDependencePerturb2} (Color online)
        Fixed-nuclei ionization yields $Y(R)$ for parallel-aligned
        H$_2$ as a function of the laser frequency $\omega$ for
        different internuclear distances $R$.  (Laser parameters as in
        Figure~\ref{fig:YieldDependencePerturb}.)}
\end{center}
\end{figure}

The dependence of the ionization yield $Y(R)$ on the internuclear
distance $R$ and the corresponding contribution
$Y(R)\left|\chi(R)\right|^2$ of internuclear distances to the
$R$-integrated ionization yield (Eq.~(\ref{eq:YieldRint})) is shown in
Fig.~\ref{fig:YieldDependencePerturb}.  In general, the ionization
yield $Y(R)$ changes many orders of magnitude with varying $R$. It
might be surprising that for $\omega = 0.2 , 0.29$, and $\geq 0.38$
a.u.\ the equilibrium distance $R_{\mathrm{eq}}=1.4$ a.u.\ practically
does not contribute at all to the total ionization yield. Of course,
large differences between the FNA and FROZ or FULL treatment are
observed for these frequencies in Fig.~\ref{fig:ComparisonFM}a. In
particular, for the breakdown of the FNA in the two-photon regime
between $0.38\ \mathrm{a.u.}  \leq \omega \leq 0.47\ \mathrm{a.u.}$,
one can see how the $R$-integrated ionization yield is dominated by
increasingly larger internuclear distances with lower and lower laser
frequency. This is compatible with the expectation stemming from the
simple picture in Fig.~\ref{fig:ComparisonFM}b.

Figure~\ref{fig:YieldDependencePerturb2} displays how the
frequency-dependence of the fixed-nuclei ionization yield $Y(R)$
changes with internuclear distance.  For increasing internuclear
distance, the threshold between $N$ and $N+1$ photon ionization shifts
to lower laser frequencies. Thus, while at the equilibrium distance
$R_{\mathrm{eq}}=1.4$ a.u.\ four-photon (three-photon) ionization
occurs at the laser frequency $\omega = 0.2$ a.u.\ ($0.29$ a.u.),
three-photon (two-photon) ionization occurs at larger internuclear
distances.  This leads to a significantly enhanced ionization yield at
larger internuclear distances and thus pronounced differences between
the fixed- and the frozen-nuclei ionization approximations (see also
Figs.~\ref{fig:ComparisonFM} and \ref{fig:YieldDependencePerturb}).
Furthermore, Fig.~\ref{fig:YieldDependencePerturb2} shows how the
previously mentioned two-photon REMPI ${X}^{1}\Sigma_g^+\ \rightarrow\
{B}^{1}\Sigma_u^+\ \rightarrow \mathrm{H}_2^+(1\sigma_g) + e^-$
requires lower and lower laser frequencies (and also strongly
increases in magnitude) with increasing internuclear distance.  The
frequency shift of the ionization thresholds and resonance frequencies
with internuclear distance can also be seen in the perturbative
ionization cross sections in Figs.~1-3 of Ref.~\cite{sfm:apal02}. As
in Ref.~\cite{sfm:awas05}, however, only the two-photon resonances are
clearly visible as peaks in the ionization yield when solving the TDSE
for short ($T=10$\,fs) pulses. Noteworthy, also the resonance due to
the second autoionizing state with ${}^{1}\Sigma_g$ symmetry belonging
to the Q(1) series, Q(1) ${}^{1}\Sigma_g(2)$, is visible at $R=2$
a.u.\ and around $\omega=0.47$ a.u.\ in
Fig.~\ref{fig:YieldDependencePerturb} (as reported in
Ref.~\cite{sfm:apal02}).  Despite the fact that the nuclear
probability density $\left|\chi(R \approx 2.0\ \mathrm{a.u.}
  )\right|^2$ is very small, this resonance still contributes
noticeably to the total ionization yield, see the second hump for
$\omega=0.47$ a.u.\ in Fig.~\ref{fig:YieldDependencePerturb}, right
panel.

So far, only a parallel alignment of the laser polarization with
respect to the molecular axis has been considered.  However, an
experiment with unaligned molecules (i.e. random alignment) can
experimentally easier be realized.  Compared to a parallel alignment,
a non-parallel alignment is computationally much more expensive.  This
is due to the broken cylindrical symmetry. In this case, not only
electronic eigenstates with ${}^{1}\Sigma_u^+$ and ${}^{1}\Sigma_u^+$
symmetry, but considerably more symmetries have to be taken into
account.  However, since a single FNA TDSE calculation for parallel
alignment in the perturbative regime is nowadays extremely fast (order
of one second), it is comparatively simple to extend the previous
study and to include the full alignment dependence together with the
$R$ dependence.

\begin{figure*}
\begin{center}
     \includegraphics[width=0.99\textwidth]{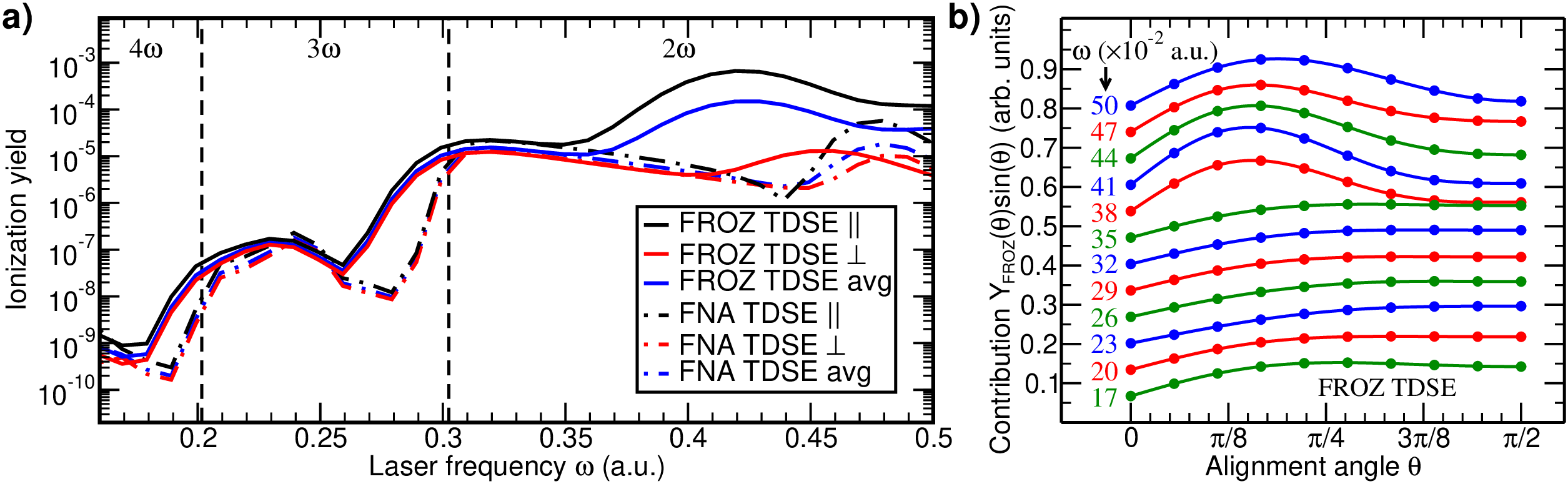}
     \caption{\label{fig:IonPertAngleDep} (Color online) a) Fixed-
       (FNA) and frozen-nuclei (FROZ) ionization yields as a function
       of the laser frequency $\omega$ for $T=10$ fs, $I=10^{12}$
       Wcm${}^{-2}$ $\cos^2$-shaped laser pulses and a parallel
       ($\parallel$), perpendicular ($\perp$) and random (avg)
       alignment of the H$_2$ molecule.  b) Contribution
       $Y_{\mathrm{FROZ}}(\theta)\sin(\theta)$ to the
       alignment-averaged frozen-nuclei ionization yield in
       Eq.~(\ref{eq:Yieldthetaint}) (for better visibility scaled and
       vertically shifted).}
\end{center}
\end{figure*}

Fig.~\ref{fig:IonPertAngleDep}a shows the FNA and the FROZ TDSE
ionization yields for a parallel, a perpendicular, and a random
alignment. The FNA TDSE behavior is similar to the behavior of the
perturbative cross sections reported in Ref.~\cite{sfm:apal02}.  For
both, FNA and FROZ TDSE, the ionization yield for a parallel alignment
is almost always higher than for a perpendicular alignment, whereas
the alignment-averaged result lies in between. The alignment
dependence of the ionization yield is rather small in the three- and
four-photon ionization regime. In contrast, the result for a
perpendicular alignment differs strongly from the parallel one in the
two-photon case, especially in the frozen-nuclei treatment. The
contributions $Y_{\mathrm{FROZ}}(\theta)\sin(\theta)$ of different
alignment angles $\theta$ to the alignment-averaged ionization yield,
see Eq.~(\ref{eq:Yieldthetaint}), are shown in
Fig.~\ref{fig:IonPertAngleDep}b. In contrast to the $R$ dependence
(Fig.~\ref{fig:YieldDependencePerturb}) the alignment dependence is
very smooth and a large portion of possible alignment angles
contributes to the total ionization. In the three- and four-photon
regime where the $\theta$ dependence of $Y_{\mathrm{FROZ}}(\theta)$ is
rather small, the geometrically preferred alignment angles $\theta
\lesssim \frac{\pi}{2}$ contribute most due to the $\sin(\theta)$
factor. In the two-photon regime, this geometrical preference competes
with a strong dependence of the ionization yield
$Y_{\mathrm{FROZ}}(\theta)$ on $\theta$ that shows the opposite
trend. In conclusion, intermediate alignment angles around $\theta
\approx \frac{\pi}{8}$ contribute most. Despite this huge alignment
dependence for two-photon ionization, the previously discussed
breakdown of the FNA is clearly seen in
Fig.~\ref{fig:IonPertAngleDep}a also for randomly aligned molecules.

\section{Intense 800\,nm laser pulses}\label{sec:800nm}
\begin{figure}
\begin{center}
     \includegraphics[width=0.46\textwidth]{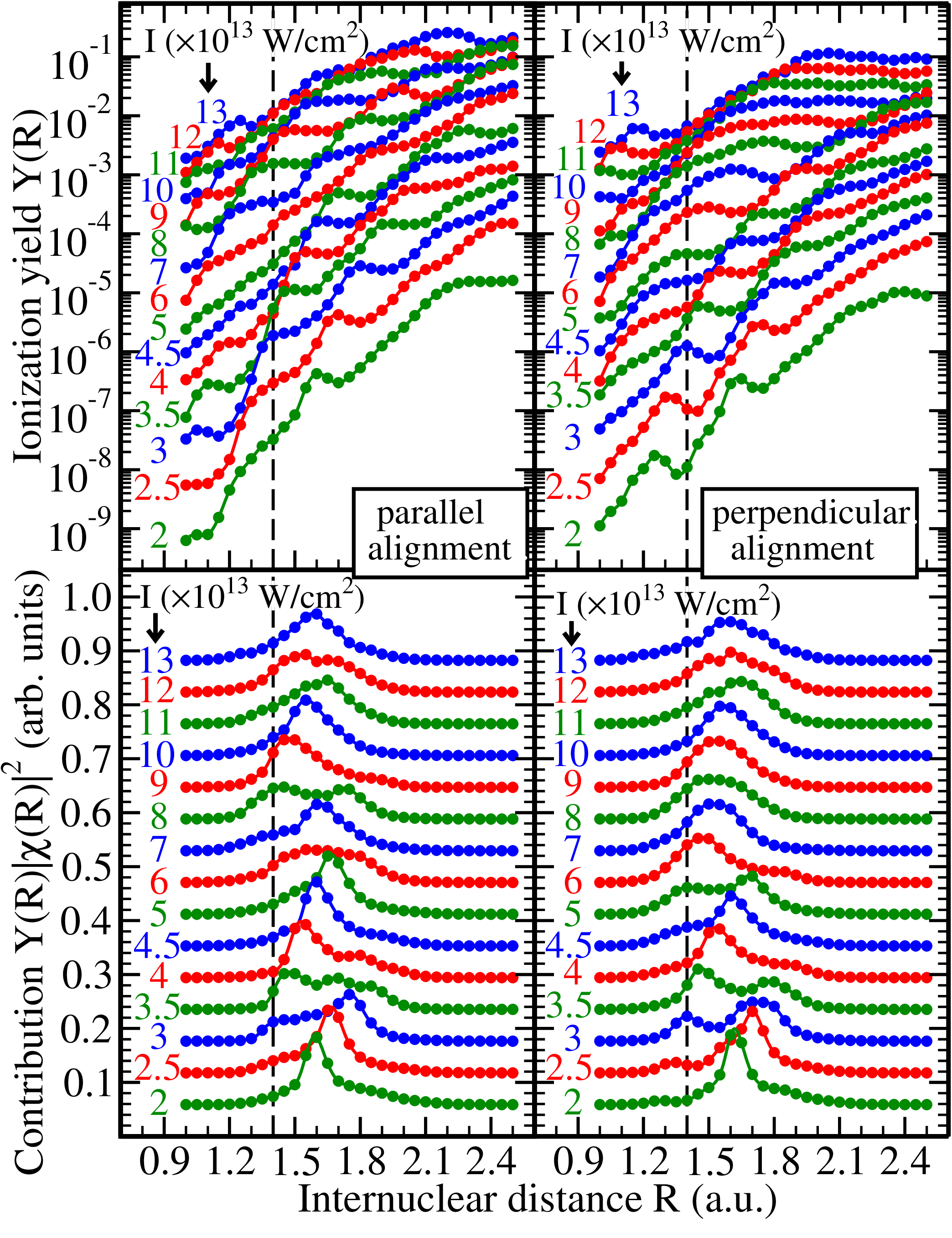}
     \caption{\label{fig:IonRdep800nm} (Color online) Ionization of
       H$_2$ molecules exposed to 20-cycle cos$^2$-shaped 800\,nm
       laser pulses and different laser peak intensities $I$.  The
       upper panel shows the fixed-nuclei ionization yields $Y(R)$,
       whereas the lower panel displays the contribution
       $Y(R)\left|\chi(R)\right|^2$ to the frozen-nuclei ionization
       yield in Eq.~(\ref{eq:YieldRint}) (scaled and vertically
       shifted).  The left (right) panel shows the result for a
       parallel (perpendicular) alignment of the molecule with respect
       to the field axis.}
       
\end{center}
\end{figure}

We investigate the ionization behavior of hydrogen molecules exposed
to intense laser pulses with the ubiquitous Ti:sapphire wavelength of
800\,nm.  The response to frequency-doubled 400\,nm laser pulses has
been studied earlier \cite{sfm:vann09}.  First, 800\,nm $\cos^2$ laser
pulses with $n_{\mathrm{c}} = 20$ cycles (FWHM of about $20$\,fs),
carrier-envelope phase $\varphi = 0$ and peak intensities $I$ varying
between $2\times 10^{13}$ to $1.3\times 10^{14}$\,W/cm$^2$ are
considered.  For this range of laser intensities, the Keldysh
parameter \cite{sfa:keld65}
\begin{eqnarray}
 \gamma = \omega \frac{\sqrt{2 I_p}}{F}
\end{eqnarray}
(with the electron binding energy $I_p(R=1.4\ \mathrm{a.u.})$ and peak
laser electric field strength $F$) varies for molecular hydrogen
between $\gamma = 0.67$ and $2.6$. This corresponds to the transition
between the quasi-static ($\gamma \ll 1$) and the multiphoton ($\gamma
\gg 1$) regime.  The dependence of the ionization yield $Y(R)$ on the
internuclear distance $R$ and the corresponding contribution
$Y(R)\left|\chi(R)\right|^2$ of internuclear distances to the
$R$-integrated ionization yield (Eq.~(\ref{eq:YieldRint})) is shown in
Fig.~\ref{fig:IonRdep800nm}.  One can see a significant increase of
the ionization yield $Y(R)$ with internuclear distance $R$, e.g.\
about 4 orders of magnitude for the intensity of
$2\times10^{13}$\,W/cm$^2$.  The $R$ dependence of $Y(R)$ is very
smooth compared to laser parameters in the perturbative regime,
Fig.~\ref{fig:YieldDependencePerturb}, and notably smoother than in
the case of 400\,nm \cite{sfm:vann09}.  This behavior was already
observed earlier for shorter 6-cycle 800\,nm pulses in
Ref.~\cite{sfm:awas06} and is expected from the quasi-static picture
in which the ionization rate depends smoothly (exponentially) on the
$R$-dependent binding energy $I_p(R)$ \cite{sfm:saen00c, sfm:saen02b}.
However, on top of this smooth behavior resonance structures can be
observed.

When comparing the result for a parallel and a perpendicular alignment
of the molecule, i.\,e.\ the left and the right panels of
Fig.~\ref{fig:IonRdep800nm}, differences occur in the resonance
behavior, especially resonance positions are shifted.  Overall one
observes that the ionization behavior is quite similar for both
alignments, i.\,e.\ the alignment dependence of the ionization yield
is not a large (orders of magnitude) effect.  The lower panel of
Fig.~\ref{fig:IonRdep800nm} shows that the main contribution
$Y(R)\left|\chi(R)\right|^2$ to the $R$-integrated ionization yields
is shifted to larger internuclear distances $R>1.4$ a.u.\ due to the
strong increase of $Y(R)$ with increasing $R$.  Nevertheless, this
effect is smaller than the previously discussed breakdown of the FNA
for two-photon ionization in Fig.~\ref{fig:YieldDependencePerturb}.

\begin{figure}
\begin{center}
     \includegraphics[width=0.49\textwidth]{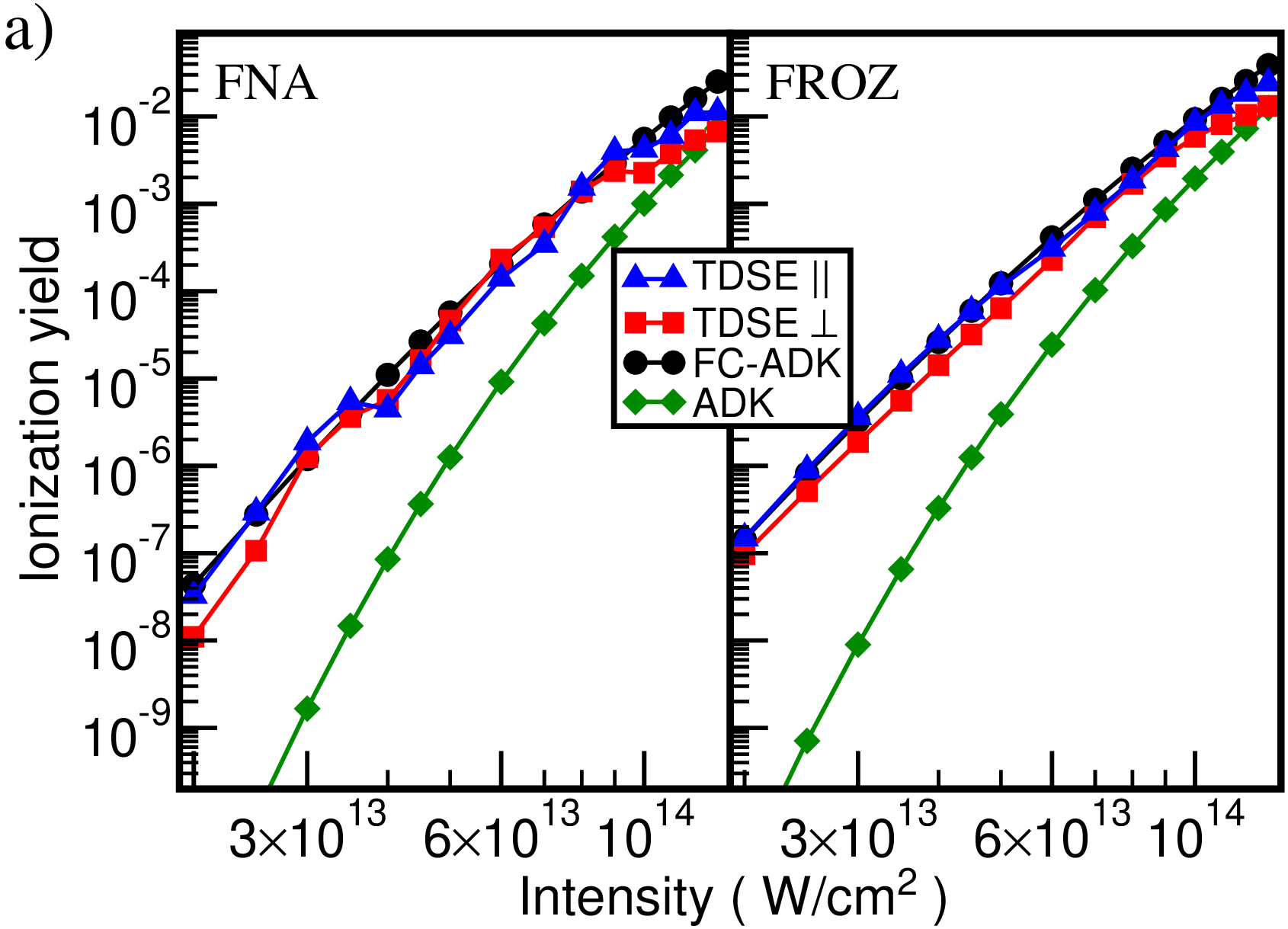}
     \includegraphics[width=0.49\textwidth]{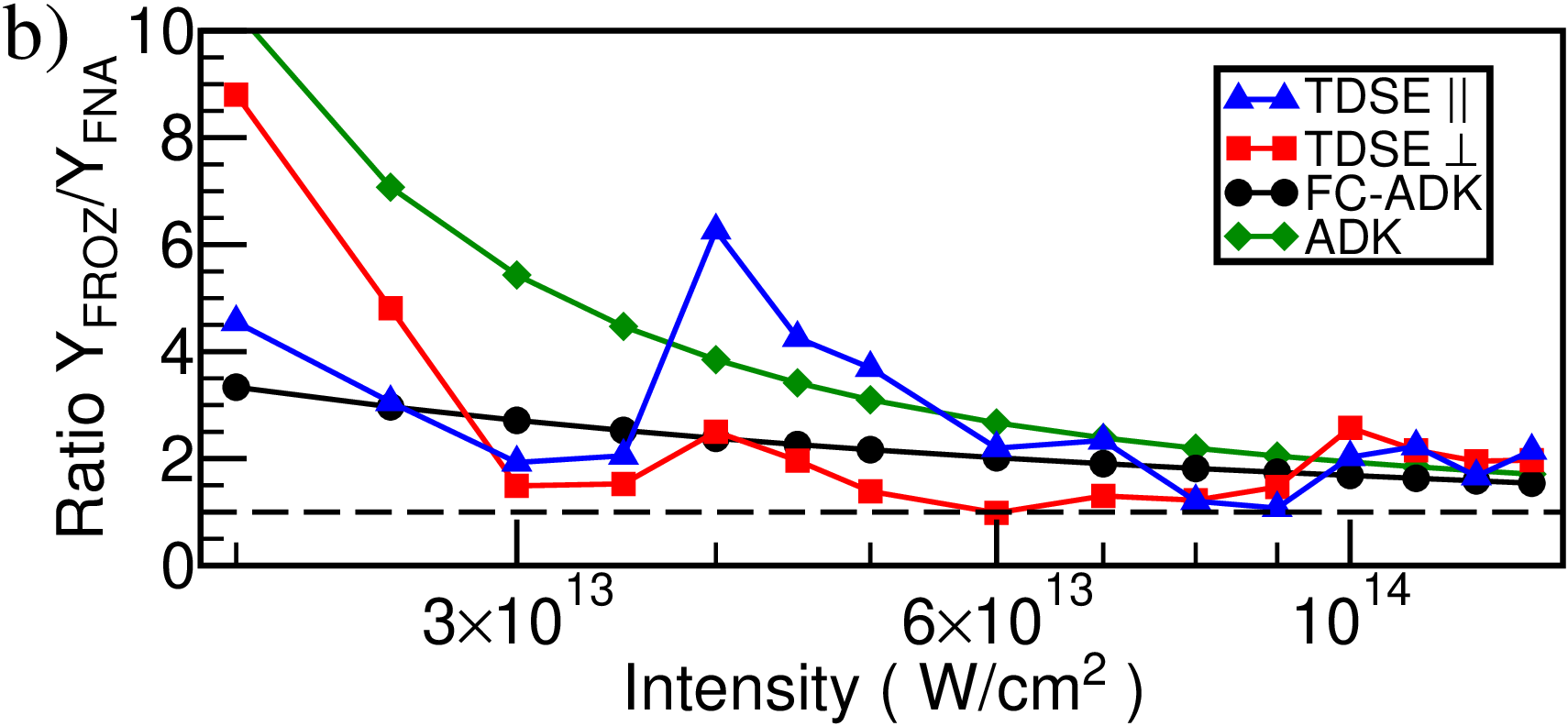}

     \caption{\label{fig:Ion800nmIdep} (Color online) Fixed- (FNA,
       left panel) and frozen-nuclei (FROZ, right panel) ionization
       yields of H$_2$ as a function of the peak intensity of a
       20-cycle cos$^2$-shaped 800\,nm laser pulse for a parallel and
       a perpendicular alignment of the molecule are compared with
       those predicted by the ADK model with (FC-ADK) and without
       (ADK) frequency correction. b) Ratio
       $Y_{\mathrm{FROZ}}/Y_{\mathrm{FNA}}$ of the ionization yields
       shown in a). The dashed horizontal line indicates
       $Y_{\mathrm{FROZ}}/Y_{\mathrm{FNA}}=1$.}
\end{center}
\end{figure}

The ionization yields can be compared to those obtained using the
Ammosov-Delone-Krainov (ADK) tunneling rates $\Gamma_\textrm{ADK}$
\cite{sfa:ammo86, sfa:ilko92}. The ion yield
\begin{equation}
  Y_\textrm{ADK}(R) =
  1-\exp\left\{-\int\!\Gamma_\textrm{ADK}[F_{\mathrm{e}}(t),I_p(R)] dt\right\}
\label{eq:Y_ADK}
\end{equation}
is obtained by integrating the tunneling rate where
$F_{\mathrm{e}}(t)$ is the envelope function of the electric field and
the integration is performed over the whole pulse duration.  For
consistency, we use the vertical binding energy $I_p(R)$ obtained from
the field-free CI calculation.  Instead of using the envelope
$F_{\mathrm{e}}(t)$ and the cycle-averaged ADK rate
$\Gamma_\textrm{ADK}$, one may also perform the integral in
Eq.~(\ref{eq:Y_ADK}) using the time-dependent electric field $F(t)$
and the static rate $\sqrt{\pi \kappa^3/(3
  F(t))}\Gamma_\textrm{ADK}[F(t),I_p(R)]$.  For the $R$-dependent
ionization yields $Y_\textrm{ADK}(R)$ shown in the following, the
relative difference when using cycle-averaged or static rates remains
below $1.3 \%$ and is thus negligible.  Noteworthy, in the here
studied transition regime with $\gamma = 0.67 - 2.6$ the validity
condition for ADK, $\gamma \ll 1$, is not (strictly) fulfilled.  The
popular ADK rates differ from the Perelomov-Popov-Terent'ev (PPT)
\cite{sfa:pere66} rates by the restriction to the quasi-static regime
$\gamma \ll 1$, the introduction of effective quantum numbers
$n^{\ast}$ and $l^{\ast}$ for non-hydrogenic atoms (or molecules), an
application of the Stirling approximation for the evaluation of
factorials, and a rearrangement of the final expression.  It is
usually assumed that the pre-exponential factor in the ionization rate
is less important than the exponential one.  Returning to the original
PPT theory \cite{sfa:pere66}, a simple correction to cycle-averaged
ADK rates $\Gamma_\textrm{ADK}$ is obtained by replacing the
exponential
\begin{eqnarray}
  \exp\left[ -\frac{2\kappa^3}{3 F_{\mathrm{e}}} \right] \rightarrow \exp\left[
  -\frac{2\kappa^3}{3 F_{\mathrm{e}}}\,\, g(\gamma) \right]
\end{eqnarray}
while leaving the prefactor unchanged. Thus, starting from Eq.~(7) in
Ref.~\cite{sfa:ilko92}, one arrives at what we call
"frequency-corrected ADK" (FC-ADK)
\begin{eqnarray}
  \Gamma_{\rm FC-ADK} = N_e \sqrt{\frac{3 F_{\mathrm{e}}}{\pi \kappa^3}
    (2/\kappa -1)} \frac{F_{\mathrm{e}}}{8\pi} \left(\frac{4 e \kappa ^
      3}{(2/\kappa -1) F_{\mathrm{e}}}\right)^{2/\kappa} \nonumber\\ \times
  \exp\left[ -\frac{2\kappa^3}{3 F_{\mathrm{e}}}\,\, g(\gamma) \right] \quad
\label{eq:IonRateCor}
\end{eqnarray}
where $e=2.718...\,$, $N_e=2$ is the number of active electrons,
$\kappa = \sqrt{2 I_p(R)}$, $\gamma = \kappa \omega /
F_{\mathrm{e}}$ and the function $g$ is defined as
\cite{sfa:pere66}
\begin{equation}
  g(\gamma) = \frac{3}{2\gamma}\biggl\{
  \Bigl(1+\frac{1}{2\gamma^2}\Bigr)\mathrm{arcsinh}\gamma -
  \frac{\sqrt{1+\gamma^2}}{2\gamma} \biggr\}\,.
\label{eq:fun_g}
\end{equation}
Noteworthy, $g(\gamma)$ is a frequency-dependent modification to the
standard ADK formula and the only $\omega$-dependent term in
Eq.~(\ref{eq:IonRateCor}).  In the limit $\gamma \ll 1$
Eq.~(\ref{eq:IonRateCor}) reduces to the standard atomic ADK rate
multiplied with the number of active electrons $N_e$.  Of course,
similar to standard ADK, FC-ADK is not well suitable for extremely
intense laser fields where, in the quasi-static length-gauge picture,
over-the-barrier ionization is possible and tunneling formulas tend to
overestimate the total ionization yield \cite{sfa:scri99,sfm:saen02b}.

Applying both ADK and FC-ADK at the equilibirum internuclear distance
leads to their predictions within the FNA, whereas the R-integration
similar to Eq.~(\ref{eq:YieldRint}) results in the predictions within
FROZ.  Fig.~\ref{fig:Ion800nmIdep}a shows the FNA and FROZ TDSE
ionization yields for parallel and for perpendicular alignment
compared to the ADK and FC-ADK results.  For both, fixed and frozen
nuclei, one observes a rather small alignment dependence, i.e.\ the
TDSE results for parallel and perpendicular alignment always agree
within a factor of $3$ (FNA) or $2$ (FROZ) with a (mostly) slightly
higher ionization yield for parallel alignment.  Assuming that the
alignment dependence relates to the symmetry of the initial state,
this result is expected since the electronic ground state of H$_2$ is
almost spherically symmetric.  For the same reason, a simple
one-electron one-center model potential \cite{sfm:vann09, sfm:vann08,
  dia:luhr08} provides a good approximation for the ionization
behavior of H$_2$.  Most interestingly, while the ionization yields
obtained with standard ADK differ from the TDSE results by several
orders of magnitude, FC-ADK and TDSE ionization yields agree
astonishingly well over the whole intensity range.  The ratio
$Y_{\mathrm{FROZ}}/Y_{\mathrm{FNA}}$ of the ionization yields in
Fig.~\ref{fig:Ion800nmIdep}a are shown in
Fig.~\ref{fig:Ion800nmIdep}b.  The ionization yield is significantly
enhanced within FROZ TDSE compared to the FNA, similar to the
breakdown of the FNA for two-photon ionization in
Fig.~\ref{fig:ComparisonFM}a. Depending on the laser intensity (and
alignment), this enhancement reaches almost one order of
magnitude. ADK and FC-ADK predict a smooth increase of the ratio with
decreasing intensity. While the overall behavior of the TDSE ratios
agree well with FC-ADK, the TDSE ratios become more and more
structured with decreasing intensity since resonance structures become
more and more pronounced (see also Fig.~\ref{fig:IonRdep800nm}).  For
example, at intensity $I=4\times 10^{13}$\,W/cm$^2$, one finds
$Y_{\mathrm{FROZ}}/Y_{\mathrm{FNA}}=6.3$ ($2.5$) in the case of a
parallel (perpendicular) alignment. At this laser intensity, a channel
closing is expected such that $13$ photons are required to overcome
the ionization threshold at $R\leq 1.35$~a.u.\ while $12$ photons are
sufficient at $R>1.35$~a.u.  This leads to a strong increase of the
ionization yield at internuclear distances which are slightly larger
than the equilibrium distance (see Fig.~\ref{fig:IonRdep800nm}) and
thus to a strongly increased ionization yield after the integration
over internuclear distances.

\begin{figure}
\begin{center}
     \includegraphics[width=0.46\textwidth]{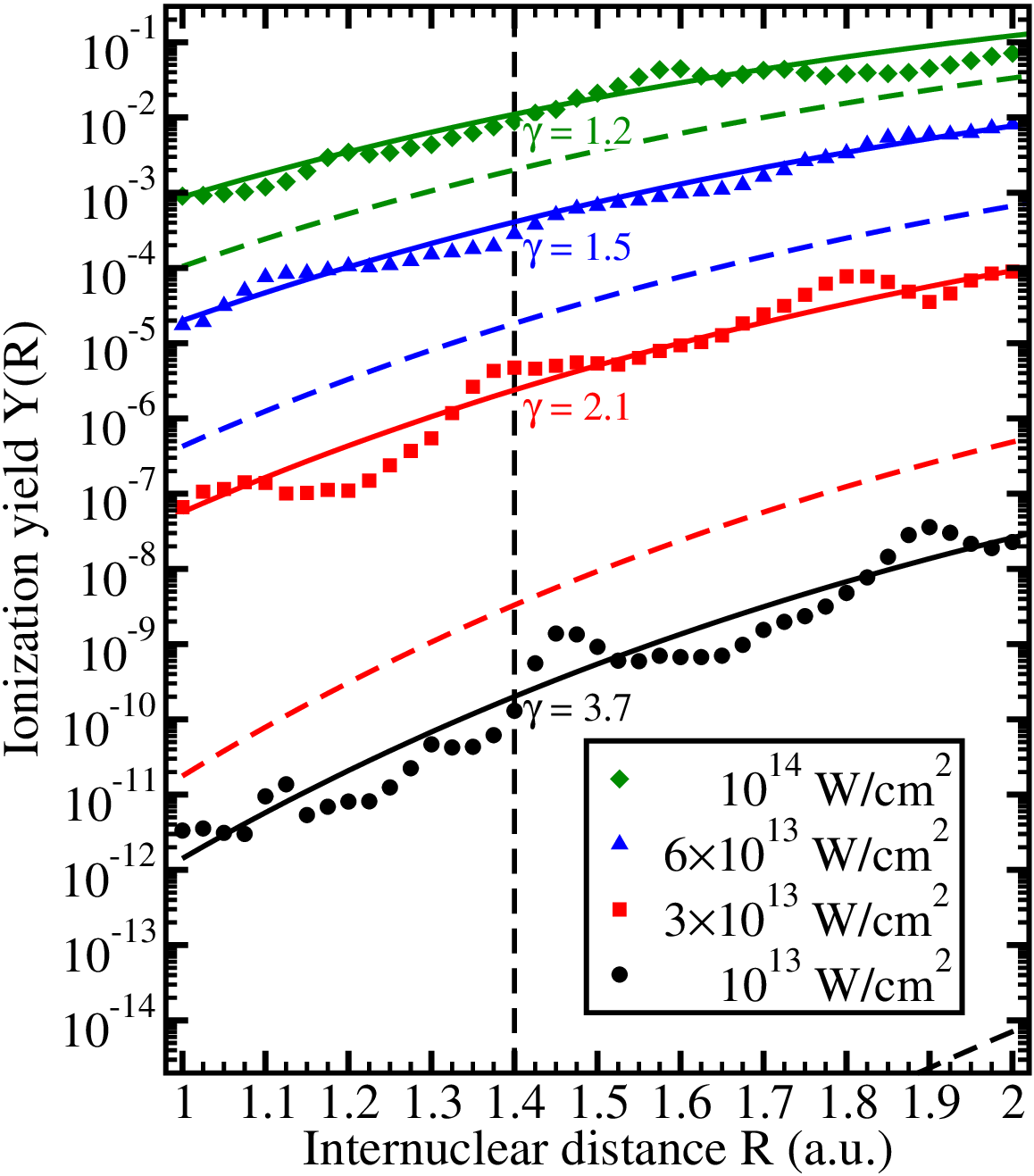}
     \caption{\label{fig:Comparison_ADK} (Color online) Ionization
       yields for a parallel-aligned H$_2$ molecule in 40-cycle
       cos$^2$-shaped 800\,nm laser pulses with different peak
       intensities are compared with those predicted using ADK (dash
       lines) and frequency-corrected ADK (solid lines) ionization
       rates. The corresponding Keldysh parameters $\gamma$ are given
       inside the graph. The dashed vertical line indicates the
       equilibrium internuclear distance $R_{\mathrm{eq}}=1.4$~a.u.\\ }
       
     \includegraphics[width=0.46\textwidth]{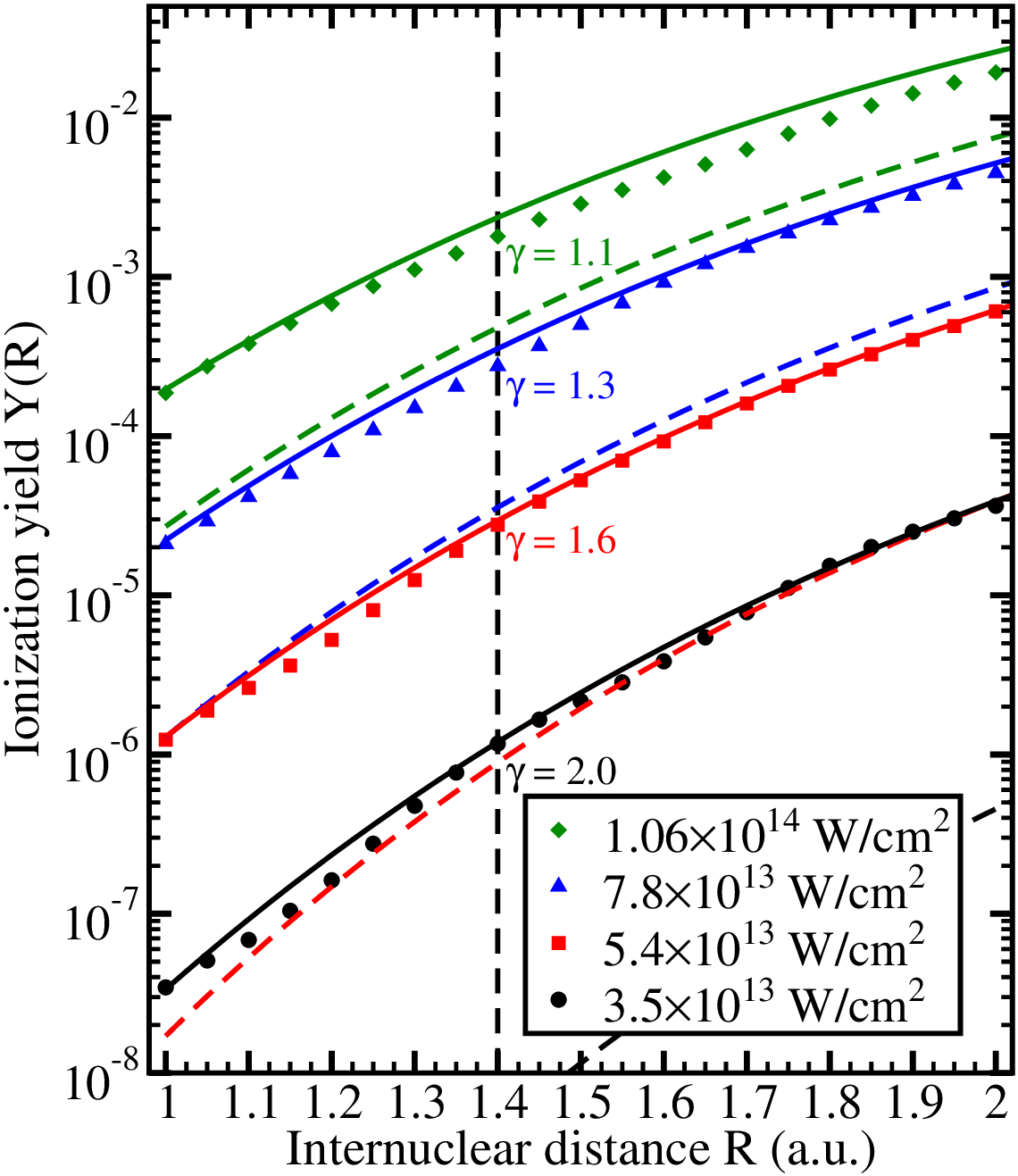}
     \caption{\label{fig:Comparison_ADK_6cyc} (Color online) As 
     Fig.~\ref{fig:Comparison_ADK}, but for 6-cycle pulses 
     and other laser peak intensities.}
\end{center}
\end{figure}

Fig.~\ref{fig:Comparison_ADK} shows the $R$-dependent ionization
yields for parallel-aligned H$_2$ exposed to 800\,nm $\cos^2$-laser
pulses with $n_{\mathrm{c}} = 40$ cycles (FWHM of about $40$ fs),
carrier-envelope phase $\varphi = 0$ and peak intensities $I$ between
$10^{13}$ to $10^{14}$\,W/cm$^2$.  One observes that the $R$
dependence becomes significantly smoother with increasing intensity,
i.e.\ multiphoton resonances are, as expected, less and less
pronounced when approaching the quasi-static regime. Comparing the
TDSE ionization yields to standard ADK, one finds that ADK
qualitatively predicts the correct $R$ dependence of the ionization
yield while it may differ quantitatively by several orders of
magnitude.  The quantitative agreement improves with increasing
intensity, i.e.\ decreasing Keldysh parameter $\gamma$.  In contrast
to standard ADK, however, FC-ADK and TDSE ionization yields agree
quantitatively surprisingly well for the whole intensity range.  It
was found already in Ref.~\cite{sfm:awas06} that it is possible to
predict the $R$ dependence of the TDSE ionization yield with ADK even
for $\gamma \gtrsim 1$, if the obtained yield $Y_\textrm{ADK}(R)$ is
multiplied with a constant prefactor.  Considerably shorter 800 nm
$\cos^2$-laser pulses with $n_{\mathrm{c}} = 6$ cycles (FWHM of about
$6$ fs), carrier-envelope phase $\varphi = 0$ and peak intensities $I$
between $3.5\times 10^{13}$ to $1.06\times 10^{14}$\,W/cm$^2$ were
investigated in Ref.~\cite{sfm:awas06}.  The prefactors needed to
match ADK to TDSE ionization yields range up to 75 for
$3.5\times10^{13}$\,W/cm$^2$ \footnote{Unfortunately, as it turns out
  the ADK rate in \cite{sfm:awas06} had an additional prefactor
  $\sqrt{\pi \kappa^3/(3 F)}$, i.\,e.\ $\sqrt{\pi \kappa^3/(3
    F)}\Gamma_{\mathrm{ADK}}$ was used instead of
  $\Gamma_{\mathrm{ADK}}$. Thus, the more correct scaling factors for
  obtaining agreement between ADK and TDSE results in
  \cite{sfm:awas06} are $75$, $14$, $5$, and $2.5$ for the laser peak
  intensities $3.5$, $5.4$, $7.8$, and
  $10.6\times~10^{13}$~W/cm${}^{-2}$, respectively.}.  We recalculate
the TDSE ionization yields for these laser pulses in order to compare
them with ADK and FC-ADK with a higher $R$ resolution.  For fully
converged results, the basis set as described in Sec.~\ref{sec:Method}
is extended by a second (long) configuration series where one electron
occupies the H$_2^+$ exited state $1\sigma_u$ while the other is
occupying one of the remaining (bound or discretized continuum)
H$_2^+$ eigenstates. The resulting TDSE ionization yields are in good
agreement with Ref.~\cite{sfm:awas06}.  The $R$-dependent ionization
yields for parallel-aligned H$_2$ exposed to these $n_{\mathrm{c}} =
6$ cycle pulses are shown in
Fig.~\ref{fig:Comparison_ADK_6cyc}. Compared to $n_{\mathrm{c}} = 40$
cycle pulses (Fig.~\ref{fig:Comparison_ADK}), resonances are much less
pronounced since a shorter pulse is broader in the frequency
domain. When compared to the TDSE, FC-ADK again predicts the correct
$R$ dependence almost quantitatively.  Large scaling factors as
required to match the behavior of ADK ionization yields to the TDSE
results are thus not required for FC-ADK.

Despite the intrinsic short-coming of FC-ADK to describe resonances,
the excellent agreement between TDSE and FC-ADK in the transition from
the quasi-static to the multiphoton regime confirms the usefulness of
the FC-ADK rates from Eq.~(\ref{eq:IonRateCor}), as shown for a range
of laser intensities and pulse durations in
Figs.~\ref{fig:Ion800nmIdep}-\ref{fig:Comparison_ADK_6cyc}.

\begin{figure}
\begin{center}

\end{center}
\end{figure}

%
\section{Conclusions}
%
The ionization behavior of molecular hydrogen exposed to high
frequency, low intensity as well as intense low-frequency (800\,nm)
laser pulses has been studied theoretically by solving the
full-dimensional time-dependent two-electron Schr\"odinger equation.
In the perturbative multiphoton ionization regime a good agreement
between our TDSE results and TDSE as well as LOPT results reported in
literature was found.  Furthermore, a surprisingly strong dependence
of the fixed-nuclei ionization yields $Y(R)$ on the internuclear
distance $R$ was found.  This effect, caused by REMPI, offers a new
explanation for the previously reported breakdown of the fixed-nuclei
approximation for two-photon ionization \cite{sfm:pala06, sfm:pala07,
  sfm:sanz07}.  The explanation, based on the frozen-nuclei
approximation, still neglects vibrational dynamics during the laser
field and considers only the extended nuclear wave function $\chi(R)$
of the initial state.  Thus, this effect is expected to be important
also for heavier molecules even though the actual laser-induced
vibrational dynamics may be negligible. Noteworthy, the frozen-nuclei
approximation is computationally much simpler than the fully coherent
treatment of electronic and nuclear motion and it provides a very
simple picture for the interpretation of results (vertical transitions
between electronic Born-Oppenheimer potentials).  The
alignment-dependence of the ionization yield turns out to be rather
small for three- and four-photon ionization. In contrast, it is very
pronounced in the two-photon regime.  Nevertheless, even for randomly
aligned molecules, the breakdown of the fixed-nuclei approximation for
two-photon ionization is clearly visible.

For intense 800\,nm laser pulses in the transition between the
multiphoton and the quasi-static regime, we found a comparably small
alignment dependence.  On the other hand, we observed a pronounced
increase of the fixed-nuclei ionization yield $Y(R)$ with increasing
internuclear distance $R$.  This increase is well understood by the
exponential dependence of the quasi-static ionization rate on the
binding energy $I_p(R)$.  The smooth $R$ dependence is superimposed by
multiphoton resonances which become less and less pronounced when
approaching the quasi-static regime.  We found that while ADK
qualitatively describes the increase of $Y(R)$ with $R$, it completely
fails quantitatively for $\gamma \gtrsim 1$ (which is outside the
validity region of ADK, $\gamma \ll 1$).  Thus, motivated by the
original PPT theory, FC-ADK was introduced as a simple modification of
the standard ADK formula.  The quantitative agreement between the
FC-ADK and the TDSE results is astonishing and manifests the
usefulness of the modified ADK formula, e.\,g.\ for the calibration of
the laser intensity in experiments.  \vspace{-0,5cm}
\begin{acknowledgments}
  The authors gratefully acknowledge financial support from the {\it
    German National Academic Foundation (Studienstiftung des deutschen
    Volkes)}, the {\it Humboldt Center for Modern Optics}, the {\it
    Fonds der Chemischen Industrie}, the {\it EU Initial Training
    Network (ITN) CORINF}, and the {\it European COST Action CM1204
    (XLIC)}. This research was supported in part by the {\it National
  Science Foundation} under Grant No.\ {\it NSF PHY11-25915}.
\end{acknowledgments}

\bibliographystyle{apsrev}

\begin{thebibliography}{37}
\expandafter\ifx\csname natexlab\endcsname\relax\def\natexlab#1{#1}\fi
\expandafter\ifx\csname bibnamefont\endcsname\relax
  \def\bibnamefont#1{#1}\fi
\expandafter\ifx\csname bibfnamefont\endcsname\relax
  \def\bibfnamefont#1{#1}\fi
\expandafter\ifx\csname citenamefont\endcsname\relax
  \def\citenamefont#1{#1}\fi
\expandafter\ifx\csname url\endcsname\relax
  \def\url#1{\texttt{#1}}\fi
\expandafter\ifx\csname urlprefix\endcsname\relax\def\urlprefix{URL }\fi
\providecommand{\bibinfo}[2]{#2}
\providecommand{\eprint}[2][]{\url{#2}}

\bibitem[{\citenamefont{Itatani et~al.}(2004)\citenamefont{Itatani, Levesque,
  Zeidler, Niikura, P{\'e}pin, Kieffer, Corkum, and Villeneuve}}]{sfm:itat04}
\bibinfo{author}{\bibfnamefont{J.}~\bibnamefont{Itatani}},
  \bibinfo{author}{\bibfnamefont{J.}~\bibnamefont{Levesque}},
  \bibinfo{author}{\bibfnamefont{D.}~\bibnamefont{Zeidler}},
  \bibinfo{author}{\bibfnamefont{H.}~\bibnamefont{Niikura}},
  \bibinfo{author}{\bibfnamefont{H.}~\bibnamefont{P{\'e}pin}},
  \bibinfo{author}{\bibfnamefont{J.~C.} \bibnamefont{Kieffer}},
  \bibinfo{author}{\bibfnamefont{P.~B.} \bibnamefont{Corkum}},
  \bibnamefont{and} \bibinfo{author}{\bibfnamefont{D.~M.}
  \bibnamefont{Villeneuve}}, \bibinfo{journal}{Nature}
  \textbf{\bibinfo{volume}{432}}, \bibinfo{pages}{867} (\bibinfo{year}{2004}).

\bibitem[{\citenamefont{Lein}(2005)}]{sfm:lein05}
\bibinfo{author}{\bibfnamefont{M.}~\bibnamefont{Lein}},
  \bibinfo{journal}{Phys.\,Rev.\,Lett.} \textbf{\bibinfo{volume}{94}},
  \bibinfo{pages}{053004} (\bibinfo{year}{2005}).

\bibitem[{\citenamefont{Baker et~al.}(2006)\citenamefont{Baker, Robinson,
  Haworth, Teng, Smith, Chiril{\u{a}}, Lein, Tisch, and Marangos}}]{sfm:bake06}
\bibinfo{author}{\bibfnamefont{S.}~\bibnamefont{Baker}},
  \bibinfo{author}{\bibfnamefont{J.~S.} \bibnamefont{Robinson}},
  \bibinfo{author}{\bibfnamefont{C.~A.} \bibnamefont{Haworth}},
  \bibinfo{author}{\bibfnamefont{H.}~\bibnamefont{Teng}},
  \bibinfo{author}{\bibfnamefont{R.~A.} \bibnamefont{Smith}},
  \bibinfo{author}{\bibfnamefont{C.~C.} \bibnamefont{Chiril{\u{a}}}},
  \bibinfo{author}{\bibfnamefont{M.}~\bibnamefont{Lein}},
  \bibinfo{author}{\bibfnamefont{J.~W.~G.} \bibnamefont{Tisch}},
  \bibnamefont{and} \bibinfo{author}{\bibfnamefont{J.~P.}
  \bibnamefont{Marangos}}, \bibinfo{journal}{Science}
  \textbf{\bibinfo{volume}{312}}, \bibinfo{pages}{424} (\bibinfo{year}{2006}).

\bibitem[{\citenamefont{Farrell et~al.}(2011)\citenamefont{Farrell, Petretti,
  F{\"o}rster, McFarland, Spector, Vanne, Decleva, Bucksbaum, Saenz, and
  G{\"u}hr}}]{sfm:farr11a}
\bibinfo{author}{\bibfnamefont{J.~P.} \bibnamefont{Farrell}},
  \bibinfo{author}{\bibfnamefont{S.}~\bibnamefont{Petretti}},
  \bibinfo{author}{\bibfnamefont{J.}~\bibnamefont{F{\"o}rster}},
  \bibinfo{author}{\bibfnamefont{B.~K.} \bibnamefont{McFarland}},
  \bibinfo{author}{\bibfnamefont{L.~S.} \bibnamefont{Spector}},
  \bibinfo{author}{\bibfnamefont{Y.~V.} \bibnamefont{Vanne}},
  \bibinfo{author}{\bibfnamefont{P.}~\bibnamefont{Decleva}},
  \bibinfo{author}{\bibfnamefont{P.~H.} \bibnamefont{Bucksbaum}},
  \bibinfo{author}{\bibfnamefont{A.}~\bibnamefont{Saenz}}, \bibnamefont{and}
  \bibinfo{author}{\bibfnamefont{M.}~\bibnamefont{G{\"u}hr}},
  \bibinfo{journal}{Phys.\,Rev.\,Lett.} \textbf{\bibinfo{volume}{107}},
  \bibinfo{pages}{083001} (\bibinfo{year}{2011}).

\bibitem[{\citenamefont{Kraus and W\"orner}(2013)}]{sfm:krau13}
\bibinfo{author}{\bibfnamefont{P.}~\bibnamefont{Kraus}} \bibnamefont{and}
  \bibinfo{author}{\bibfnamefont{H.}~\bibnamefont{W\"orner}},
  \bibinfo{journal}{Chem.\,Phys.\,Chem.} \textbf{\bibinfo{volume}{14}},
  \bibinfo{pages}{1445} (\bibinfo{year}{2013}).

\bibitem[{\citenamefont{F\"orster and Saenz}(2013)}]{sfm:foer13}
\bibinfo{author}{\bibfnamefont{J.}~\bibnamefont{F\"orster}} \bibnamefont{and}
  \bibinfo{author}{\bibfnamefont{A.}~\bibnamefont{Saenz}},
  \bibinfo{journal}{Chem.\,Phys.\,Chem.} \textbf{\bibinfo{volume}{14}},
  \bibinfo{pages}{1438} (\bibinfo{year}{2013}).

\bibitem[{\citenamefont{W\"orner et~al.}(2010)\citenamefont{W\"orner, Bertrand,
  Kartashov, Corkum, and Villeneuve}}]{sfm:worn10c}
\bibinfo{author}{\bibfnamefont{H.~J.} \bibnamefont{W\"orner}},
  \bibinfo{author}{\bibfnamefont{J.~B.} \bibnamefont{Bertrand}},
  \bibinfo{author}{\bibfnamefont{D.~V.} \bibnamefont{Kartashov}},
  \bibinfo{author}{\bibfnamefont{P.~B.} \bibnamefont{Corkum}},
  \bibnamefont{and} \bibinfo{author}{\bibfnamefont{D.~M.}
  \bibnamefont{Villeneuve}}, \bibinfo{journal}{Nature}
  \textbf{\bibinfo{volume}{466}}, \bibinfo{pages}{604} (\bibinfo{year}{2010}).

\bibitem[{\citenamefont{W{\"o}rner et~al.}(2011)\citenamefont{W{\"o}rner,
  Bertrand, Fabre, Higuet, Ruf, Dubrouil, Patchkovskii, Spanner, Mairesse,
  Blanchet et~al.}}]{sfm:woer11}
\bibinfo{author}{\bibfnamefont{H.~J.} \bibnamefont{W{\"o}rner}},
  \bibinfo{author}{\bibfnamefont{J.~B.} \bibnamefont{Bertrand}},
  \bibinfo{author}{\bibfnamefont{B.}~\bibnamefont{Fabre}},
  \bibinfo{author}{\bibfnamefont{J.}~\bibnamefont{Higuet}},
  \bibinfo{author}{\bibfnamefont{H.}~\bibnamefont{Ruf}},
  \bibinfo{author}{\bibfnamefont{A.}~\bibnamefont{Dubrouil}},
  \bibinfo{author}{\bibfnamefont{S.}~\bibnamefont{Patchkovskii}},
  \bibinfo{author}{\bibfnamefont{M.}~\bibnamefont{Spanner}},
  \bibinfo{author}{\bibfnamefont{Y.}~\bibnamefont{Mairesse}},
  \bibinfo{author}{\bibfnamefont{V.}~\bibnamefont{Blanchet}},
  \bibnamefont{et~al.}, \bibinfo{journal}{Science}
  \textbf{\bibinfo{volume}{334}}, \bibinfo{pages}{208} (\bibinfo{year}{2011}).

\bibitem[{\citenamefont{Meckel et~al.}(2008)\citenamefont{Meckel, Comtois,
  Zeidler, Staudte, Pavi{\v{c}}i{\'c}, Bandulet, P{\'e}pin, Kieffer,
  D{\"o}rner, Villeneuve et~al.}}]{sfm:meck08}
\bibinfo{author}{\bibfnamefont{M.}~\bibnamefont{Meckel}},
  \bibinfo{author}{\bibfnamefont{D.}~\bibnamefont{Comtois}},
  \bibinfo{author}{\bibfnamefont{D.}~\bibnamefont{Zeidler}},
  \bibinfo{author}{\bibfnamefont{A.}~\bibnamefont{Staudte}},
  \bibinfo{author}{\bibfnamefont{D.}~\bibnamefont{Pavi{\v{c}}i{\'c}}},
  \bibinfo{author}{\bibfnamefont{H.~C.} \bibnamefont{Bandulet}},
  \bibinfo{author}{\bibfnamefont{H.}~\bibnamefont{P{\'e}pin}},
  \bibinfo{author}{\bibfnamefont{J.~C.} \bibnamefont{Kieffer}},
  \bibinfo{author}{\bibfnamefont{R.}~\bibnamefont{D{\"o}rner}},
  \bibinfo{author}{\bibfnamefont{D.~M.} \bibnamefont{Villeneuve}},
  \bibnamefont{et~al.}, \bibinfo{journal}{Science}
  \textbf{\bibinfo{volume}{320}}, \bibinfo{pages}{1478} (\bibinfo{year}{2008}).

\bibitem[{\citenamefont{Petretti et~al.}(2010)\citenamefont{Petretti, Vanne,
  Saenz, Castro, and Decleva}}]{sfm:petr10a}
\bibinfo{author}{\bibfnamefont{S.}~\bibnamefont{Petretti}},
  \bibinfo{author}{\bibfnamefont{Y.~V.} \bibnamefont{Vanne}},
  \bibinfo{author}{\bibfnamefont{A.}~\bibnamefont{Saenz}},
  \bibinfo{author}{\bibfnamefont{A.}~\bibnamefont{Castro}}, \bibnamefont{and}
  \bibinfo{author}{\bibfnamefont{P.}~\bibnamefont{Decleva}},
  \bibinfo{journal}{Phys.\,Rev.\,Lett.} \textbf{\bibinfo{volume}{104}},
  \bibinfo{pages}{223001} (\bibinfo{year}{2010}).

\bibitem[{\citenamefont{Goll et~al.}(2006)\citenamefont{Goll, Wunner, and
  Saenz}}]{sfm:goll06}
\bibinfo{author}{\bibfnamefont{E.}~\bibnamefont{Goll}},
  \bibinfo{author}{\bibfnamefont{G.}~\bibnamefont{Wunner}}, \bibnamefont{and}
  \bibinfo{author}{\bibfnamefont{A.}~\bibnamefont{Saenz}},
  \bibinfo{journal}{Phys.\,Rev.\,Lett.} \textbf{\bibinfo{volume}{97}},
  \bibinfo{pages}{103003} (\bibinfo{year}{2006}).

\bibitem[{\citenamefont{Ergler et~al.}(2006)\citenamefont{Ergler, Feuerstein,
  Rudenko, Zrost, Schr{\"o}ter, Moshammer, and Ullrich}}]{sfm:ergl06}
\bibinfo{author}{\bibfnamefont{T.}~\bibnamefont{Ergler}},
  \bibinfo{author}{\bibfnamefont{B.}~\bibnamefont{Feuerstein}},
  \bibinfo{author}{\bibfnamefont{A.}~\bibnamefont{Rudenko}},
  \bibinfo{author}{\bibfnamefont{K.}~\bibnamefont{Zrost}},
  \bibinfo{author}{\bibfnamefont{C.~D.} \bibnamefont{Schr{\"o}ter}},
  \bibinfo{author}{\bibfnamefont{R.}~\bibnamefont{Moshammer}},
  \bibnamefont{and} \bibinfo{author}{\bibfnamefont{J.}~\bibnamefont{Ullrich}},
  \bibinfo{journal}{Phys.\,Rev.\,Lett.} \textbf{\bibinfo{volume}{97}},
  \bibinfo{pages}{103004} (\bibinfo{year}{2006}).

\bibitem[{\citenamefont{Fang and Gibson}(2008)}]{sfm:fang08a}
\bibinfo{author}{\bibfnamefont{L.}~\bibnamefont{Fang}} \bibnamefont{and}
  \bibinfo{author}{\bibfnamefont{G.~N.} \bibnamefont{Gibson}},
  \bibinfo{journal}{Phys.\,Rev.\,Lett.} \textbf{\bibinfo{volume}{100}},
  \bibinfo{pages}{103003} (\bibinfo{year}{2008}).

\bibitem[{\citenamefont{Mart{\'\i}n}(1999)}]{dia:mart99}
\bibinfo{author}{\bibfnamefont{F.}~\bibnamefont{Mart{\'\i}n}},
  \bibinfo{journal}{J.\,Phys.\,B} \textbf{\bibinfo{volume}{32}},
  \bibinfo{pages}{R197} (\bibinfo{year}{1999}).

\bibitem[{\citenamefont{Apalategui and Saenz}(2002)}]{sfm:apal02}
\bibinfo{author}{\bibfnamefont{A.}~\bibnamefont{Apalategui}} \bibnamefont{and}
  \bibinfo{author}{\bibfnamefont{A.}~\bibnamefont{Saenz}},
  \bibinfo{journal}{J.\,Phys.\,B} \textbf{\bibinfo{volume}{35}},
  \bibinfo{pages}{1909} (\bibinfo{year}{2002}).

\bibitem[{\citenamefont{Harumiya et~al.}(2000)\citenamefont{Harumiya, Kawata,
  Kono, and Fujimura}}]{sfm:haru00}
\bibinfo{author}{\bibfnamefont{K.}~\bibnamefont{Harumiya}},
  \bibinfo{author}{\bibfnamefont{I.}~\bibnamefont{Kawata}},
  \bibinfo{author}{\bibfnamefont{H.}~\bibnamefont{Kono}}, \bibnamefont{and}
  \bibinfo{author}{\bibfnamefont{Y.}~\bibnamefont{Fujimura}},
  \bibinfo{journal}{J.\,Chem.\,Phys.} \textbf{\bibinfo{volume}{113}},
  \bibinfo{pages}{8953} (\bibinfo{year}{2000}).

\bibitem[{\citenamefont{Awasthi et~al.}(2005)\citenamefont{Awasthi, Vanne, and
  Saenz}}]{sfm:awas05}
\bibinfo{author}{\bibfnamefont{M.}~\bibnamefont{Awasthi}},
  \bibinfo{author}{\bibfnamefont{Y.~V.} \bibnamefont{Vanne}}, \bibnamefont{and}
  \bibinfo{author}{\bibfnamefont{A.}~\bibnamefont{Saenz}},
  \bibinfo{journal}{J.\,Phys.\,B} \textbf{\bibinfo{volume}{38}},
  \bibinfo{pages}{3973} (\bibinfo{year}{2005}).

\bibitem[{\citenamefont{Awasthi and Saenz}(2006)}]{sfm:awas06}
\bibinfo{author}{\bibfnamefont{M.}~\bibnamefont{Awasthi}} \bibnamefont{and}
  \bibinfo{author}{\bibfnamefont{A.}~\bibnamefont{Saenz}},
  \bibinfo{journal}{J.\,Phys.\,B} \textbf{\bibinfo{volume}{39}},
  \bibinfo{pages}{S\,389} (\bibinfo{year}{2006}).

\bibitem[{\citenamefont{Vanne and Saenz}(2008)}]{sfm:vann08}
\bibinfo{author}{\bibfnamefont{Y.~V.} \bibnamefont{Vanne}} \bibnamefont{and}
  \bibinfo{author}{\bibfnamefont{A.}~\bibnamefont{Saenz}},
  \bibinfo{journal}{J.\,Mod.\,Opt.} \textbf{\bibinfo{volume}{55}},
  \bibinfo{pages}{2665} (\bibinfo{year}{2008}).

\bibitem[{\citenamefont{Vanne and Saenz}(2009)}]{sfm:vann09}
\bibinfo{author}{\bibfnamefont{Y.~V.} \bibnamefont{Vanne}} \bibnamefont{and}
  \bibinfo{author}{\bibfnamefont{A.}~\bibnamefont{Saenz}},
  \bibinfo{journal}{Phys.\,Rev.\,A} \textbf{\bibinfo{volume}{80}},
  \bibinfo{pages}{053422} (\bibinfo{year}{2009}).

\bibitem[{\citenamefont{Vanne and Saenz}(2010)}]{sfm:vann10}
\bibinfo{author}{\bibfnamefont{Y.~V.} \bibnamefont{Vanne}} \bibnamefont{and}
  \bibinfo{author}{\bibfnamefont{A.}~\bibnamefont{Saenz}},
  \bibinfo{journal}{Phys.\,Rev.\,A} \textbf{\bibinfo{volume}{82}},
  \bibinfo{pages}{011403} (\bibinfo{year}{2010}).

\bibitem[{\citenamefont{Palacios et~al.}(2006)\citenamefont{Palacios, Bachau,
  and Mart{\'i}n}}]{sfm:pala06}
\bibinfo{author}{\bibfnamefont{A.}~\bibnamefont{Palacios}},
  \bibinfo{author}{\bibfnamefont{H.}~\bibnamefont{Bachau}}, \bibnamefont{and}
  \bibinfo{author}{\bibfnamefont{F.}~\bibnamefont{Mart{\'i}n}},
  \bibinfo{journal}{Phys.\,Rev.\,Lett.} \textbf{\bibinfo{volume}{96}},
  \bibinfo{pages}{143001} (\bibinfo{year}{2006}).

\bibitem[{\citenamefont{Palacios et~al.}(2007)\citenamefont{Palacios, Bachau,
  and Mart{\'i}n}}]{sfm:pala07}
\bibinfo{author}{\bibfnamefont{A.}~\bibnamefont{Palacios}},
  \bibinfo{author}{\bibfnamefont{H.}~\bibnamefont{Bachau}}, \bibnamefont{and}
  \bibinfo{author}{\bibfnamefont{F.}~\bibnamefont{Mart{\'i}n}},
  \bibinfo{journal}{Phys.\,Rev.\,A} \textbf{\bibinfo{volume}{75}},
  \bibinfo{pages}{013408} (\bibinfo{year}{2007}).

\bibitem[{\citenamefont{Sanz-Vicario et~al.}(2007)\citenamefont{Sanz-Vicario,
  Palacios, Cardona, Bachau, and Mart{\'i}n}}]{sfm:sanz07}
\bibinfo{author}{\bibfnamefont{J.~L.} \bibnamefont{Sanz-Vicario}},
  \bibinfo{author}{\bibfnamefont{A.}~\bibnamefont{Palacios}},
  \bibinfo{author}{\bibfnamefont{J.~C.} \bibnamefont{Cardona}},
  \bibinfo{author}{\bibfnamefont{H.}~\bibnamefont{Bachau}}, \bibnamefont{and}
  \bibinfo{author}{\bibfnamefont{F.}~\bibnamefont{Mart{\'i}n}},
  \bibinfo{journal}{J.\,Electr.\,Spectros.\,Relat.\,Phenom.}
  \textbf{\bibinfo{volume}{161}}, \bibinfo{pages}{182} (\bibinfo{year}{2007}).

\bibitem[{\citenamefont{Rivi\`{e}re et~al.}(2012)\citenamefont{Rivi\`{e}re,
  Silva, and Mart{\'{i}}n}}]{sfm:rivi12}
\bibinfo{author}{\bibfnamefont{P.}~\bibnamefont{Rivi\`{e}re}},
  \bibinfo{author}{\bibfnamefont{R.~E.~F.} \bibnamefont{Silva}},
  \bibnamefont{and}
  \bibinfo{author}{\bibfnamefont{F.}~\bibnamefont{Mart{\'{i}}n}},
  \bibinfo{journal}{J.\,Phys.\,Chem.\,A} \textbf{\bibinfo{volume}{116}},
  \bibinfo{pages}{11304} (\bibinfo{year}{2012}).

\bibitem[{\citenamefont{Silva et~al.}(2012)\citenamefont{Silva, Rivi\`{e}re,
  and Mart{\'{i}}n}}]{sfm:silv12}
\bibinfo{author}{\bibfnamefont{R.~E.~F.} \bibnamefont{Silva}},
  \bibinfo{author}{\bibfnamefont{P.}~\bibnamefont{Rivi\`{e}re}},
  \bibnamefont{and}
  \bibinfo{author}{\bibfnamefont{F.}~\bibnamefont{Mart{\'{i}}n}},
  \bibinfo{journal}{Phys.\,Rev.\,A} \textbf{\bibinfo{volume}{85}},
  \bibinfo{pages}{063414} (\bibinfo{year}{2012}).

\bibitem[{\citenamefont{Dehghanian et~al.}(2010)\citenamefont{Dehghanian,
  Bandrauk, and Kamta}}]{sfm:dehg10}
\bibinfo{author}{\bibfnamefont{E.}~\bibnamefont{Dehghanian}},
  \bibinfo{author}{\bibfnamefont{A.~D.} \bibnamefont{Bandrauk}},
  \bibnamefont{and} \bibinfo{author}{\bibfnamefont{G.~L.} \bibnamefont{Kamta}},
  \bibinfo{journal}{Phys.\,Rev.\,A} \textbf{\bibinfo{volume}{81}},
  \bibinfo{pages}{061403} (\bibinfo{year}{2010}).

\bibitem[{\citenamefont{Ammosov et~al.}(1986)\citenamefont{Ammosov, Delone, and
  Krainov}}]{sfa:ammo86}
\bibinfo{author}{\bibfnamefont{M.~V.} \bibnamefont{Ammosov}},
  \bibinfo{author}{\bibfnamefont{N.~B.} \bibnamefont{Delone}},
  \bibnamefont{and} \bibinfo{author}{\bibfnamefont{V.~P.}
  \bibnamefont{Krainov}}, \bibinfo{journal}{Sov.\,Phys.\,JETP}
  \textbf{\bibinfo{volume}{64}}, \bibinfo{pages}{1191} (\bibinfo{year}{1986}).

\bibitem[{\citenamefont{Perelomov et~al.}(1966)\citenamefont{Perelomov, Popov,
  and Terent'ev}}]{sfa:pere66}
\bibinfo{author}{\bibfnamefont{A.~M.} \bibnamefont{Perelomov}},
  \bibinfo{author}{\bibfnamefont{V.~S.} \bibnamefont{Popov}}, \bibnamefont{and}
  \bibinfo{author}{\bibfnamefont{M.~V.} \bibnamefont{Terent'ev}},
  \bibinfo{journal}{Sov.\,Phys.\,JETP} \textbf{\bibinfo{volume}{23}},
  \bibinfo{pages}{924} (\bibinfo{year}{1966}).

\bibitem[{\citenamefont{Vanne and Saenz}(2004)}]{dia:vann04}
\bibinfo{author}{\bibfnamefont{Y.~V.} \bibnamefont{Vanne}} \bibnamefont{and}
  \bibinfo{author}{\bibfnamefont{A.}~\bibnamefont{Saenz}},
  \bibinfo{journal}{J.\,Phys.\,B} \textbf{\bibinfo{volume}{37}},
  \bibinfo{pages}{4101} (\bibinfo{year}{2004}).

\bibitem[{\citenamefont{Keldysh}(1965)}]{sfa:keld65}
\bibinfo{author}{\bibfnamefont{L.~V.} \bibnamefont{Keldysh}},
  \bibinfo{journal}{Sov.\,Phys.\,JETP} \textbf{\bibinfo{volume}{20}},
  \bibinfo{pages}{1307} (\bibinfo{year}{1965}).

\bibitem[{\citenamefont{Saenz}(2000)}]{sfm:saen00c}
\bibinfo{author}{\bibfnamefont{A.}~\bibnamefont{Saenz}},
  \bibinfo{journal}{J.\,Phys.\,B} \textbf{\bibinfo{volume}{33}},
  \bibinfo{pages}{4365} (\bibinfo{year}{2000}).

\bibitem[{\citenamefont{Saenz}(2002{\natexlab{a}})}]{sfm:saen02b}
\bibinfo{author}{\bibfnamefont{A.}~\bibnamefont{Saenz}},
  \bibinfo{journal}{Phys.\,Rev.\,A} \textbf{\bibinfo{volume}{66}},
  \bibinfo{pages}{063408} (\bibinfo{year}{2002}{\natexlab{a}}).

\bibitem[{\citenamefont{Ilkov et~al.}(1992)\citenamefont{Ilkov, Decker, and
  Chin}}]{sfa:ilko92}
\bibinfo{author}{\bibfnamefont{F.~A.} \bibnamefont{Ilkov}},
  \bibinfo{author}{\bibfnamefont{J.~E.} \bibnamefont{Decker}},
  \bibnamefont{and} \bibinfo{author}{\bibfnamefont{S.~L.} \bibnamefont{Chin}},
  \bibinfo{journal}{J.\,Phys.\,B} \textbf{\bibinfo{volume}{25}},
  \bibinfo{pages}{4005} (\bibinfo{year}{1992}).

\bibitem[{\citenamefont{Scrinzi et~al.}(1999)\citenamefont{Scrinzi, Geissler,
  and Brabec}}]{sfa:scri99}
\bibinfo{author}{\bibfnamefont{A.}~\bibnamefont{Scrinzi}},
  \bibinfo{author}{\bibfnamefont{M.}~\bibnamefont{Geissler}}, \bibnamefont{and}
  \bibinfo{author}{\bibfnamefont{T.}~\bibnamefont{Brabec}},
  \bibinfo{journal}{Phys.\,Rev.\,Lett.} \textbf{\bibinfo{volume}{83}},
  \bibinfo{pages}{706} (\bibinfo{year}{1999}).

\bibitem[{\citenamefont{L\"uhr et~al.}(2008)\citenamefont{L\"uhr, Vanne, and
  Saenz}}]{dia:luhr08}
\bibinfo{author}{\bibfnamefont{A.}~\bibnamefont{L\"uhr}},
  \bibinfo{author}{\bibfnamefont{Y.~V.} \bibnamefont{Vanne}}, \bibnamefont{and}
  \bibinfo{author}{\bibfnamefont{A.}~\bibnamefont{Saenz}},
  \bibinfo{journal}{Phys.\,Rev.\,A} \textbf{\bibinfo{volume}{78}},
  \bibinfo{pages}{042510} (\bibinfo{year}{2008}).

\bibitem[{\citenamefont{Saenz}(2002{\natexlab{b}})}]{sfm:saen02a}
\bibinfo{author}{\bibfnamefont{A.}~\bibnamefont{Saenz}},
  \bibinfo{journal}{Phys.\,Rev.\,A} \textbf{\bibinfo{volume}{66}},
  \bibinfo{pages}{063407} (\bibinfo{year}{2002}{\natexlab{b}}).

\end{thebibliography}

\end{document}